\documentclass[zpreprint,zbstpl]{./zeus_paper}
\usepackage[english]{babel}
\usepackage{graphicx}
\chardef\usc=95
\chardef\til=126
\catcode`\@=11 
\DeclareRobustCommand\xdotspace{\futurelet\@let@token\@xdotspace}
\def\@xdotspace{%
  \ifx\@let@token.\else
  \ifx\@let@token\bgroup.\else
  \ifx\@let@token\egroup.\else
  \ifx\@let@token\/.\else
  \ifx\@let@token\ .\else
  \ifx\@let@token~.\else
  \ifx\@let@token!.\else
  \ifx\@let@token,.\else
  \ifx\@let@token:.\else
  \ifx\@let@token;.\else
  \ifx\@let@token?.\else
  \ifx\@let@token/.\else
  \ifx\@let@token'.\else
  \ifx\@let@token).\else
  \ifx\@let@token-.\else
  \ifx\@let@token\@xobeysp.\else
  \ifx\@let@token\space.\else
  \ifx\@let@token\@sptoken.\else
   .\space
   \fi\fi\fi\fi\fi\fi\fi\fi\fi\fi\fi\fi\fi\fi\fi\fi\fi\fi}
\catcode`\@=12 

\newcommand{\stru}[2]{%
   \relax\ifmmode\hbox{\vrule height#1 depth#2 width0pt}%
   \else\vrule height#1 depth#2 width0pt\fi}

\newcommand{\Ronum}[1]{\uppercase\expandafter{\romannumeral#1}}
\newcommand{\ronum}[1]{\expandafter{\romannumeral#1}}
\DeclareRobustCommand{\LaTeXZ}{%
  \LaTeX\kern-.05em4\kern-.1em
  {\raisebox{-0.2ex}{$\scriptstyle\text{ZEUS}$}}\xspace}

\DeclareMathAlphabet{\mathbf}{OT1}{cmr}{bx}{sl}
\newcommand{\eVdist}{\kern-0.06667em}

\newcommand{\gev}{{\,\text{Ge}\eVdist\text{V\/}}}

\newcommand{\met}{\,\text{m}}
\newcommand{\nm}{\,\text{nm}}

\newcommand{\mm}{\,\text{mm}}
\newcommand{\cm}{\,\text{cm}}
\newcommand{\km}{\,\text{km}}

\newcommand{\Hz}{\,\text{Hz}}
\newcommand{\kHz}{\,\text{kHz}}
\newcommand{\MHz}{\,\text{MHz}}
\newcommand{\GHz}{\,\text{GHz}}

\newcommand{\rad}{\,\text{rad}}

\newcommand{\mrad}{\,\text{mrad}}

\newcommand{\slashfrac}[2]{%
  \raisebox{0.5ex}{\ensuremath #1}\kern-0.12em/\kern-0.08em
  \raisebox{-.8ex}{\ensuremath #2}}

\newcommand{\sqr}[3]{%
    {\vcenter{\hrule height.#3ex\hbox{\vrule width.#2ex height#1ex
     \kern#1ex\vrule width.#3ex}\hrule height.#2ex}}}

\catcode`\@=11 
\newcommand{\parenbar}{\mathpalette\p@renb@r}
\def\p@renb@r#1#2{\vbox{%
  \ifx#1\scriptscriptstyle \dimen@.7em\dimen@ii.2em\else
  \ifx#1\scriptstyle \dimen@.8em\dimen@ii.25em\else
  \dimen@1em\dimen@ii.4em\fi\fi \offinterlineskip
  \ialign{\hfill##\hfill\cr
    \vbox{\hrule width\dimen@ii}\cr
    \noalign{\vskip-.3ex}%
    \hbox to\dimen@{$\mathchar300\hfil\mathchar301$}\cr
    \noalign{\vskip-.3ex}%
    $#1#2$\cr}}}
\catcode`\@=12 

\newcommand{\pom}{\mbox{\scriptsize\it I\hspace{-0.5ex}P}}

\newcommand{\IP}{{\rm I$\kern-0.01667em$P}\xspace}
\newcommand{\JB}{{\rm JB}}

\newcommand{\jet}{{\rm jet}}

\mathchardef\qsm=63
\mathchardef\pls=43
\mathchardef\mns=512
\mathchardef\plm=518
\mathchardef\eql=61
\mathchardef\smallleft=300
\mathchardef\smallright=301
\mathchardef\les=316
\mathchardef\gre=318
\mathchardef\leq=532
\mathchardef\grq=533

\catcode`\@=11 
\newcounter{pict@width}
\newcounter{pict@height}
\newlength{\pict@scale}
\newcommand{\psfigadd}[4]{%
\setcounter{pict@width}{1*\ratio{#2+\pict@scale/2}{\pict@scale}}
\setcounter{pict@height}{1*\ratio{#3+\pict@scale/2}{\pict@scale}}
\setlength{\unitlength}{\pict@scale}
\hbox to #2{\hspace{-\fill}\begin{picture}(\thepict@width,\thepict@height)
\put(0,0){\psfig{figure=#1,width=#2,height=#3,clip=}}
\SetScale{0.283466457}
\SetWidth{1.763889}
{#4}
\end{picture}}
}
\newcounter{pict@widthfst}
\newcounter{pict@widthscd}
\newcounter{pict@widthtot}
\newcommand{\psfigaddtwo}[7]{%
\setcounter{pict@widthfst}{1*\ratio{#2+\pict@scale/2}{\pict@scale}}
\setcounter{pict@widthscd}{1*\ratio{#2+#4+\pict@scale/2}{\pict@scale}}
\setcounter{pict@widthtot}{1*\ratio{#2+#4+#6+\pict@scale/2}{\pict@scale}}
\setcounter{pict@height}{1*\ratio{#3+\pict@scale/2}{\pict@scale}}
\setlength{\unitlength}{\pict@scale}
\hbox{\hspace{-\fill}\begin{picture}(\thepict@widthtot,\thepict@height)
\put(0,0){\psfig{figure=#1,width=#2,height=#3,clip=}}
\put(\thepict@widthscd,0){\psfig{figure=#5,width=#6,height=#3,clip=}}
\SetScale{0.283466457}
\SetWidth{1.763889}
{#7}
\end{picture}}
}
\newcommand{\psfigror}[4]{%
\setcounter{pict@width}{1*\ratio{#2+\pict@scale/2}{\pict@scale}}
\setcounter{pict@height}{1*\ratio{#3+\pict@scale/2}{\pict@scale}}
\setlength{\unitlength}{\pict@scale}
\hbox{\begin{picture}(\thepict@width,\thepict@height)
\put(0,\thepict@height){\psfig{figure=#1,width=#3,height=#2,clip=,angle=270}}
\SetScale{0.283466457}
\SetWidth{1.763889}
{#4}
\end{picture}}
}
\newcommand{\psfigrol}[4]{%
\setcounter{pict@width}{1*\ratio{#2+\pict@scale/2}{\pict@scale}}
\setcounter{pict@height}{1*\ratio{#3+\pict@scale/2}{\pict@scale}}
\setlength{\unitlength}{\pict@scale}
\hbox{\begin{picture}(\thepict@width,\thepict@height)
\put(0,0){\psfig{figure=#1,width=#3,height=#2,clip=,angle=90}}
\SetScale{0.283466457}
\SetWidth{1.763889}
{#4}
\end{picture}}
}
\catcode`\@=12 
\newlength\listtextwidth

\catcode`\@=11 
\newlength{\@tabfninsert}
\newlength{\@tabfnwidth}
\newcommand{\tabfootnote}[2]{%
  \setlength{\@tabfninsert}{0.8em}
  \setlength{\@tabfnwidth}{\textwidth}
  \addtolength{\@tabfnwidth}{-\@tabfninsert}
  \addtolength{\@tabfnwidth}{-0.4em}
  \noindent\makebox[\@tabfninsert][r]{\footnotesize$^{#1}$\hfil}\hfill%
  \parbox[t]{\@tabfnwidth}{\footnotesize #2\hfill}}
\catcode`\@=12 

\newcommand{\ZcoosysAA}{%
The ZEUS coordinate system is a right-handed Cartesian system, with the $Z$
axis pointing in the proton beam direction, referred to as the ``forward
direction'', and the $X$ axis pointing  towards the center of HERA.
The coordinate origin is at the nominal interaction point.\xspace}
\newcommand{\ZcoosysAE}{%
The ZEUS coordinate system is a right-handed Cartesian system, with the $Z$
axis pointing in the proton beam direction, referred to as the ``forward
direction'', and the $X$ axis pointing  towards the centre of HERA.
The coordinate origin is at the nominal interaction point.\xspace}
\newcommand{\ZcoosysBA}{%
The ZEUS coordinate system is a right-handed Cartesian system, with the $Z$ 
axis pointing in the nominal proton beam direction, referred to as the ``forward
direction'', and the $X$ axis pointing  towards the center of HERA.
The coordinate origin is at the centre of the CTD.\xspace}
\newcommand{\ZcoosysBE}{%
The ZEUS coordinate system is a right-handed Cartesian system, with the $Z$ 
axis pointing in the nominal proton beam direction, referred to as the ``forward
direction'', and the $X$ axis pointing  towards the centre of HERA.
The coordinate origin is at the centre of the CTD.\xspace}
\newcommand{\ZcoosysCE}{%
The ZEUS coordinate system is a right-handed Cartesian system, with the $Z$ 
axis pointing in the nominal proton beam direction, referred to as the ``forward
direction'', and the $X$ axis pointing  towards the centre of HERA.
The coordinate origin is at the centre of the central tracking detector.\xspace}
\newcommand{\ZpsrapA}{%
The pseudorapidity is defined as $\eta=-\ln\left(\tan\frac{\theta}{2}\right)$,
where the polar angle, $\theta$, is measured with respect to the
$Z$ axis.\xspace}
\newcommand{\ZpsrapB}{%
The pseudorapidity is defined as $\eta=-\ln\left(\tan\frac{\theta}{2}\right)$, 
where the polar angle, $\theta$, is measured with respect to the
$Z$ axis. The azimuthal angle, $\phi$, is
measured with respect to the $X$ axis.\xspace}

\newcommand{\ZcoosysfnCEeta}{\footnote{\ZcoosysCE\ZpsrapA}}

\newcommand{\Zsttdesc}[1]{%
The STT consisted of 48 sectors of two different sizes. Each sector
contained 192 (small sector) or 264 (large sector) straws of diameter
\unit[7.5]{\mm} arranged into 3 layers. The sectors were trapezoidal in shape
and each subtended an azimuthal angle of $60\degree$ -- 6 sectors
formed a so-called superlayer. A particle passing through the complete
detector traversed 8 superlayers, which were rotated around the beam
direction at angles of $30\degree$ or $15\degree$ to each other. The STT
covered the polar-angle region $5\degree < \theta < 23\degree$.
}
\def\citeCTD{{\cite{%
nim:a279:290,*npps:b32:181,*nim:a338:254%
}}\xspace}
\def\citeMVD{{\cite{%
nim:a581:656%
}}\xspace}
\def\citeCAL{{\cite{%
nim:a309:77,*nim:a309:101,*nim:a321:356,*nim:a336:23%
}}\xspace}

\newcommand{\PYTHIA}{\textsc{Pythia}\xspace}

\newcommand{\RAPGAP}{\textsc{Rapgap}\xspace}
\newcommand{\DJANGOH}{\textsc{Djangoh}\xspace}
\newcommand{\ARIADNE}{\textsc{Ariadne}\xspace}
\newcommand{\GRAPE}{\textsc{Grape-Compton}\xspace}

\newcommand{\ET}{\ensuremath{E_{T}}\xspace}
\newcommand{\ETjet}{\ensuremath{E_{T}^{\mathrm{jet}}}\xspace}
\newcommand{\Ejet}{\ensuremath{E^{\mathrm{jet}}}\xspace}
\newcommand{\pzjet}{\ensuremath{p_Z^{\mathrm{jet}}}\xspace}

\newcommand{\Egam}{\ensuremath{E^{\gamma}}\xspace}
\newcommand{\pzgam}{\ensuremath{p_Z^{\gamma}}\xspace}
\newcommand{\ETgam}{\ensuremath{E_{T}^{\gamma}}\xspace}
\newcommand{\etagam}{\ensuremath{\eta^{\gamma}}\xspace}
\newcommand{\etajet}{\ensuremath{\eta^{\mathrm{jet}}}\xspace}
\newcommand{\etamax}{\ensuremath{\eta_{\mathrm{max}}}\xspace}
\newcommand{\delphi}{\ensuremath{\Delta\phi}\xspace}

\newcommand{\Deleta}{\ensuremath{\Delta\eta}\xspace}

\newcommand{\xgamm}{\ensuremath{x_{\gamma}^{\mathrm{meas}}}\xspace}
\newcommand{\xpom}{\ensuremath{x_{\pom}}\xspace}
\newcommand{\zpom}{\ensuremath{z_{\pom}^{\mathrm{meas}}}\xspace}
\newcommand{\Mx}{\ensuremath{M_{X}}\xspace}
\newcommand{\Zacknowledge}{%
We appreciate the contributions to the construction, maintenance and operation of
the ZEUS detector made by many people who are not listed as authors. The
HERA machine group and the DESY computing staff are especially
acknowledged for their success in providing excellent operation of the
collider and the data-analysis environment. We thank the DESY
directorate for their strong support and encouragement.}

\begin{document}
\prepnum{DESY-17-077}

\title{
Studies of the diffractive photoproduction of isolated photons at HERA
}                                                       
                    
\author{ZEUS Collaboration}
\abstract{
The photoproduction of isolated photons has been measured in 
diffractive events recorded by the ZEUS detector at HERA.  Cross
sections are evaluated in the photon transverse-energy and
pseudorapidity ranges
\mbox{$5 < E_T^{\gamma} < 15\,\gev$} and \mbox{$-0.7 < \eta^{\gamma} <
0.9$,} inclusively and also with a jet with transverse energy and
pseudorapidity in the ranges $4 < \ETjet < 35\,\gev$ and $-1.5 <
\etajet < 1.8$, using a total integrated electron--proton luminosity of 456
$\mathrm{pb}^{-1}$. A number of kinematic variables were studied and
compared to predictions from the \RAPGAP\ Monte Carlo model.  An
excess of data is observed above the \RAPGAP\ predictions for $\zpom >
0.9$, where \zpom\ is the fraction of the longitudinal momentum of the
colourless ``Pomeron'' exchange that is transferred to the photon--jet
final state, giving evidence for direct Pomeron interactions.

}
\makezeustitle

%
%
%
%

                                                   %
\begin{center}
{                      \Large  The ZEUS Collaboration              }
\end{center}

{\small\raggedright


H.~Abramowicz$^{24, o}$, 
I.~Abt$^{19}$, 
L.~Adamczyk$^{7}$, 
M.~Adamus$^{30}$, 
S.~Antonelli$^{1}$, 
V.~Aushev$^{16}$, 
Y.~Aushev$^{16}$, 
O.~Behnke$^{9}$, 
U.~Behrens$^{9}$, 
A.~Bertolin$^{21}$
% ...
\newpage


{\setlength{\parskip}{0.4em}
\makebox[3ex]{$^{1}$}
\begin{minipage}[t]{14cm}
{\it INFN Bologna, Bologna, Italy}~$^{A}$

\end{minipage}

\makebox[3ex]{$^{2}$}
\begin{minipage}[t]{14cm}
{\it Physikalisches Institut der Universit\"at Bonn,
Bonn, Germany}~$^{B}$

\end{minipage}

\makebox[3ex]{$^{3}$}
\begin{minipage}[t]{14cm}
{\it Panjab University, Department of Physics, Chandigarh, India}

\end{minipage}

\makebox[3ex]{$^{4}$}
\begin{minipage}[t]{14cm}
{\it Calabria University,
Physics Department and INFN, Cosenza, Italy}~$^{A}$

\end{minipage}

\makebox[3ex]{$^{5}$}
\begin{minipage}[t]{14cm}
{\it National Centre for Particle Physics, Universiti Malaya, 50603 Kuala Lumpur, Malaysia}~$^{C}$

\end{minipage}

\makebox[3ex]{$^{6}$}
\begin{minipage}[t]{14cm}
{\it The Henryk Niewodniczanski Institute of Nuclear Physics, Polish Academy of \\
Sciences, Krakow, Poland}

\end{minipage}

\makebox[3ex]{$^{7}$}
\begin{minipage}[t]{14cm}
{\it AGH University of Science and Technology, Faculty of Physics and Applied Computer
Science, Krakow, Poland}

\end{minipage}

\makebox[3ex]{$^{8}$}
\begin{minipage}[t]{14cm}
{\it Department of Physics, Jagellonian University, Krakow, Poland}

\end{minipage}

\makebox[3ex]{$^{9}$}
\begin{minipage}[t]{14cm}
{\it Deutsches Elektronen-Synchrotron DESY, Hamburg, Germany}

\end{minipage}

\makebox[3ex]{$^{10}$}
\begin{minipage}[t]{14cm}
{\it Deutsches Elektronen-Synchrotron DESY, Zeuthen, Germany}

\end{minipage}

\makebox[3ex]{$^{11}$}
\begin{minipage}[t]{14cm}
{\it School of Physics and Astronomy, University of Glasgow,
Glasgow, United Kingdom}~$^{D}$

\end{minipage}

\makebox[3ex]{$^{12}$}
\begin{minipage}[t]{14cm}
{\it Hamburg University, Institute of Experimental Physics, Hamburg,
Germany}~$^{E}$

\end{minipage}

\makebox[3ex]{$^{13}$}
\begin{minipage}[t]{14cm}
{\it Institute of Particle and Nuclear Studies, KEK,
Tsukuba, Japan}~$^{F}$

\end{minipage}

\makebox[3ex]{$^{14}$}
\begin{minipage}[t]{14cm}
{\it Institute of Physics and Technology of Ministry of Education and
Science of Kazakhstan, Almaty, Kazakhstan}

\end{minipage}

\makebox[3ex]{$^{15}$}
\begin{minipage}[t]{14cm}
{\it Institute for Nuclear Research, National Academy of Sciences, Kyiv, Ukraine}

\end{minipage}

\makebox[3ex]{$^{16}$}
\begin{minipage}[t]{14cm}
{\it Department of Nuclear Physics, National Taras Shevchenko University of Kyiv, Kyiv, Ukraine}

\end{minipage}

\makebox[3ex]{$^{17}$}
\begin{minipage}[t]{14cm}
{\it Meiji Gakuin University, Faculty of General Education,
Yokohama, Japan}~$^{F}$

\end{minipage}

\makebox[3ex]{$^{18}$}
\begin{minipage}[t]{14cm}
{\it Lomonosov Moscow State University, Skobeltsyn Institute of Nuclear Physics,
Moscow, Russia}~$^{G}$

\end{minipage}

\makebox[3ex]{$^{19}$}
\begin{minipage}[t]{14cm}
{\it Max-Planck-Institut f\"ur Physik, M\"unchen, Germany}

\end{minipage}

\makebox[3ex]{$^{20}$}
\begin{minipage}[t]{14cm}
{\it Department of Physics, University of Oxford,
Oxford, United Kingdom}~$^{D}$

\end{minipage}

\makebox[3ex]{$^{21}$}
\begin{minipage}[t]{14cm}
{\it INFN Padova, Padova, Italy}~$^{A}$

\end{minipage}

\makebox[3ex]{$^{22}$}
\begin{minipage}[t]{14cm}
{\it Dipartimento di Fisica e Astronomia dell' Universit\`a and INFN,
Padova, Italy}~$^{A}$

\end{minipage}

\makebox[3ex]{$^{23}$}
\begin{minipage}[t]{14cm}
{\it Polytechnic University, Tokyo, Japan}~$^{F}$

\end{minipage}

\makebox[3ex]{$^{24}$}
\begin{minipage}[t]{14cm}
{\it Raymond and Beverly Sackler Faculty of Exact Sciences, School of Physics, \\
Tel Aviv University, Tel Aviv, Israel}~$^{H}$

\end{minipage}

\makebox[3ex]{$^{25}$}
\begin{minipage}[t]{14cm}
{\it Department of Physics, Tokyo Institute of Technology,
Tokyo, Japan}~$^{F}$

\end{minipage}

\makebox[3ex]{$^{26}$}
\begin{minipage}[t]{14cm}
{\it Universit\`a di Torino and INFN, Torino, Italy}~$^{A}$

\end{minipage}

\makebox[3ex]{$^{27}$}
\begin{minipage}[t]{14cm}
{\it Universit\`a del Piemonte Orientale, Novara, and INFN, Torino,
Italy}~$^{A}$

\end{minipage}

\makebox[3ex]{$^{28}$}
\begin{minipage}[t]{14cm}
{\it Physics and Astronomy Department, University College London,
London, United Kingdom}~$^{D}$

\end{minipage}

\makebox[3ex]{$^{29}$}
\begin{minipage}[t]{14cm}
{\it Faculty of Physics, University of Warsaw, Warsaw, Poland}

\end{minipage}

\makebox[3ex]{$^{30}$}
\begin{minipage}[t]{14cm}
{\it National Centre for Nuclear Research, Warsaw, Poland}

\end{minipage}

\makebox[3ex]{$^{31}$}
\begin{minipage}[t]{14cm}
{\it Department of Particle Physics and Astrophysics, Weizmann
Institute, Rehovot, Israel}

\end{minipage}

\makebox[3ex]{$^{32}$}
\begin{minipage}[t]{14cm}
{\it Department of Physics, York University, Ontario, Canada M3J 1P3}~$^{I}$

\end{minipage}

}

\vspace{3em}


{\setlength{\parskip}{0.4em}\raggedright
\makebox[3ex]{$^{ A}$}
\begin{minipage}[t]{14cm}
 supported by the Italian National Institute for Nuclear Physics (INFN) \
\end{minipage}

\makebox[3ex]{$^{ B}$}
\begin{minipage}[t]{14cm}
 supported by the German Federal Ministry for Education and Research (BMBF), under
 contract No.\ 05 H09PDF\
\end{minipage}

\makebox[3ex]{$^{ C}$}
\begin{minipage}[t]{14cm}
 supported by HIR grant UM.C/625/1/HIR/149 and UMRG grants RU006-2013, RP012A-13AFR and RP012B-13AFR from
 Universiti Malaya, and ERGS grant ER004-2012A from the Ministry of Education, Malaysia\
\end{minipage}

\makebox[3ex]{$^{ D}$}
\begin{minipage}[t]{14cm}
 supported by the Science and Technology Facilities Council, UK\
\end{minipage}

\makebox[3ex]{$^{ E}$}
\begin{minipage}[t]{14cm}
 supported by the German Federal Ministry for Education and Research (BMBF), under
 contract No.\ 05h09GUF, and the SFB 676 of the Deutsche Forschungsgemeinschaft (DFG) \
\end{minipage}

\makebox[3ex]{$^{ F}$}
\begin{minipage}[t]{14cm}
 supported by the Japanese Ministry of Education, Culture, Sports, Science and Technology
 (MEXT) and its grants for Scientific Research\
\end{minipage}

\makebox[3ex]{$^{ G}$}
\begin{minipage}[t]{14cm}
 partially supported by RF Presidential grant NSh-7989.2016.2\
\end{minipage}

\makebox[3ex]{$^{ H}$}
\begin{minipage}[t]{14cm}
 supported by the Israel Science Foundation\
\end{minipage}

\makebox[3ex]{$^{ I}$}
\begin{minipage}[t]{14cm}
 supported by the Natural Sciences and Engineering Research Council of Canada (NSERC) \
\end{minipage}

}

\pagebreak[4]
{\setlength{\parskip}{0.4em}


\makebox[3ex]{$^{ a}$}
\begin{minipage}[t]{14cm}
now at INFN Roma, Italy\
\end{minipage}

\makebox[3ex]{$^{ b}$}
\begin{minipage}[t]{14cm}
now at Sant Longowal Institute of Engineering and Technology, Longowal, Punjab, India\
\end{minipage}

\makebox[3ex]{$^{ c}$}
\begin{minipage}[t]{14cm}
also at Agensi Nuklear Malaysia, 43000 Kajang, Bangi, Malaysia\
\end{minipage}

\makebox[3ex]{$^{ d}$}
\begin{minipage}[t]{14cm}
now at Rockefeller University, New York, NY 10065, USA\
\end{minipage}

\makebox[3ex]{$^{ e}$}
\begin{minipage}[t]{14cm}
now at University of Liverpool, United Kingdom\
\end{minipage}

\makebox[3ex]{$^{ f}$}
\begin{minipage}[t]{14cm}
now at Tel Aviv University, Israel\
\end{minipage}

\makebox[3ex]{$^{ g}$}
\begin{minipage}[t]{14cm}
now at Physikalisches Institut, Universit\"{a}t Heidelberg, Germany\
\end{minipage}

\makebox[3ex]{$^{ h}$}
\begin{minipage}[t]{14cm}
supported by the Alexander von Humboldt Foundation\
\end{minipage}

\makebox[3ex]{$^{ i}$}
\begin{minipage}[t]{14cm}
Alexander von Humboldt Professor; also at DESY and University of Oxford\
\end{minipage}

\makebox[3ex]{$^{ j}$}
\begin{minipage}[t]{14cm}
also at DESY\
\end{minipage}

\makebox[3ex]{$^{ k}$}
\begin{minipage}[t]{14cm}
now at European X-ray Free-Electron Laser facility GmbH, Hamburg, Germany\
\end{minipage}

\makebox[3ex]{$^{ l}$}
\begin{minipage}[t]{14cm}
now at Kobe University, Japan\
\end{minipage}

\makebox[3ex]{$^{ m}$}
\begin{minipage}[t]{14cm}
now at RWTH Aachen, Germany\
\end{minipage}

\makebox[3ex]{$^{ n}$}
\begin{minipage}[t]{14cm}
also at University of Bristol, United Kingdom\
\end{minipage}

\makebox[3ex]{$^{ o}$}
\begin{minipage}[t]{14cm}
also at Max Planck Institute for Physics, Munich, Germany, External Scientific Member\
\end{minipage}

\makebox[3ex]{$^{ p}$}
\begin{minipage}[t]{14cm}
now at University of Bath, United Kingdom\
\end{minipage}

\makebox[3ex]{$^{ q}$}
\begin{minipage}[t]{14cm}
also supported by DESY and the Alexander von Humboldt Foundation\
\end{minipage}

\makebox[3ex]{$^{ r}$}
\begin{minipage}[t]{14cm}
also at \L\'{o}d\'{z} University, Poland\
\end{minipage}

\makebox[3ex]{$^{ s}$}
\begin{minipage}[t]{14cm}
member of \L\'{o}d\'{z} University, Poland\
\end{minipage}

\makebox[3ex]{$^{ t}$}
\begin{minipage}[t]{14cm}
now at Polish Air Force Academy in Deblin\
\end{minipage}

}

}

\pagenumbering{arabic} 
\pagestyle{plain}

\section{Introduction}
\label{sec-int}

Diffractive interactions are a distinctive class of hadronic
interactions in which the scattering of the incoming particle is
mediated by an exchanged object carrying no quantum numbers, commonly
referred to as the Pomeron.  Such processes are typically
characterised by a forward nucleon or nucleonic state that is
separated by a gap in rapidity from the hadronic final state produced
in the central region of the event. At the HERA $ep$ collider,
diffractive processes have been studied both in photoproduction and in
deep inelastic scattering (DIS), photoproduction processes being those
in which the exchanged photon is quasi-real.  The virtuality $Q^2$ of
the exchanged photon is typically much smaller than 1 \gev$^2$ in
photoproduction processes, which constitute a large majority of the
$ep$ collisions. Events with $Q^2 > 1$\gev$^2$  are conventionally regarded as
DIS.

The physical nature of the Pomeron is not fully established within
quantum chromodynamics (QCD), and a number of models have been
proposed~\cite{mrw,twogluons,rapgap}.  In an approach originated by
Ingelman and Schlein~\cite{ingelman}, the Pomeron is taken to be
a hadron-like object that contains quarks and gluons. The Pomeron parton
density functions (PDFs) can be evaluated from fits
to DIS data~\cite{epjc:48:715}.  In an alternative
approach~\cite{twogluons}, the Pomeron is equivalent to the exchange
of two gluons.

\begin{figure}

\begin{center}
\hspace*{0.02\linewidth}\includegraphics[width=0.95\linewidth]{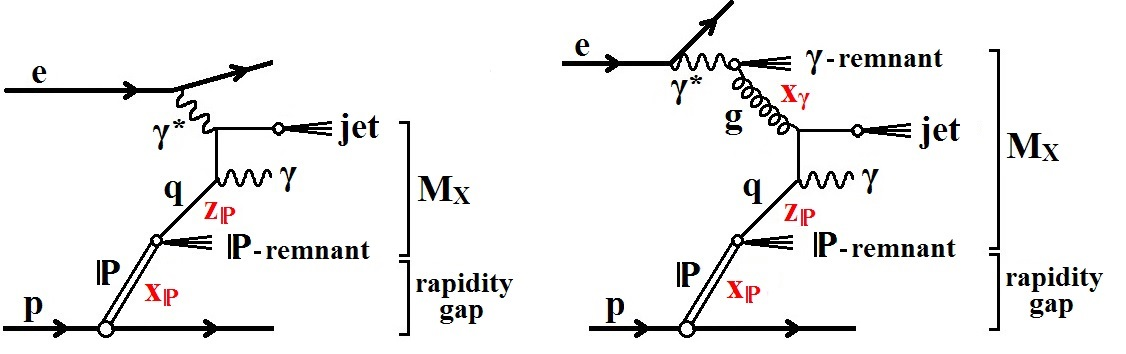}
~\\[1mm]
(a)\hspace{7.5cm}
(b)\\[5mm]
\hspace*{0.02\linewidth}\includegraphics[width=0.45\linewidth]{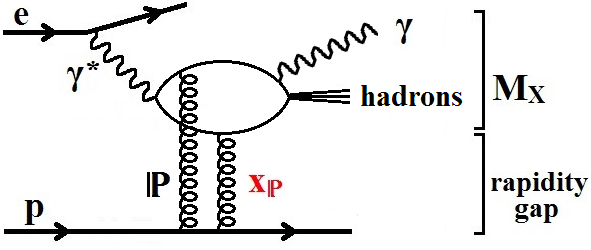}
~\\[1mm]
(c)
\end{center}
\vspace*{0mm}
\caption{Examples of diagrams for the diffractive 
production of a prompt photon and a jet in $ep$ scattering from (a)
direct (b) resolved photons, interacting with a resolved Pomeron.  The
variables are described in Section~\ref{kin-q}. (c) Example of an interaction
between a direct photon and a direct Pomeron~\protect\cite{mrw,zp:c65:657}. 
}
\label{fig1}
\end{figure}

 The photon--Pomeron interaction can take place through processes in
which the photon or Pomeron acts as a source of quarks and gluons,
which then take part in the QCD scatter (resolved processes) and
processes in which the photon or Pomeron interacts as a whole (direct
processes).  There are thus in principle four different types of
process that may be experimentally studied: a direct or resolved
photon interacting with a direct or resolved Pomeron. Examples of
these processes are illustrated in Fig.~\ref{fig1}.  Direct Pomeron
processes are not included in the Ingelman--Schlein model, but are
taken into account in other approaches~\cite{zp:c65:657}.  Within the
Ingelman--Schlein framework, it is normally assumed that a Pomeron
with a universal set of PDFs is emitted, making allowance for QCD
evolution effects.  In the H1 DIS analysis~\cite{epjc:48:715}, the
results of which are used here, the Pomeron PDFs are dominated by
gluons in most regions of parameter space, but a significant quark
content is also present. If the factorisation hypothesis holds, the
same parton structure would be valid both in direct photoproduction
processes and in DIS, although in resolved photon processes,
absorptive effects may be present~\cite{epj:c21:521,kandk,Collins}.

Several studies of diffractive dijet events in photoproduction and DIS 
have been carried out at
HERA \cite{epj:c6:421,epj:c55:177,epj:c70:15,epj:c51:549,epj:c5:41,jhep:05:056}. 
The present paper gives measurements of diffractive events in which
a hard isolated ``prompt'' photon is detected in the central region of
the ZEUS detector and may be accompanied by one or more jets.  Such
processes, while rare, are interesting for a number of reasons.  The
four different types of direct and resolved processes can be
identified, in particular direct Pomeron interactions.  The prompt
photon must originate from a charged parton, and its observation
therefore demonstrates the presence either of a quark in the Pomeron
or of higher-order processes in which both the Pomeron and the
incident photon couple to quarks. This contrasts with diffractive
dijet production, which is mainly sensitive to the gluon content of
the Pomeron.

Hard photons are also produced in ``fragmentation processes''
in which a photon is radiated within a jet. Such processes can be
suppressed by requiring the observed hard photon to be isolated from other
particles in the event.

The H1 collaboration previously measured inclusive diffractive
high-energy prompt photons as a function of their transverse momentum, but
in a different kinematic region from the present
work~\cite{plb:672:219}.  Analyses of isolated hard photons in
non-diffractive photoproduction have been presented by the ZEUS and H1
collaborations
\cite{pl:b730:293,pl:b413:201,pl:b472:175,pl:b511:19,epj:c49:511,epj:c38:437,epj:c66:17},
as well as in DIS~\cite{pl:b595:86,epj:c54:371,pl:b687:16,pl:b715:88}.

\section{The ZEUS detector} 
\label{sec-exp} 
The analysis presented here is based on two data samples corresponding
to integrated luminosities of 82 and $374\,\mathrm{pb}^{-1}$, taken
during the years 1998--2000 and 2004--2007, respectively, with the
ZEUS detector at HERA. These are referred to as HERA-I and HERA-II
samples.  During these periods, HERA ran with electron and positron
beams\footnote{Hereafter, ``electron'' refers to both electrons and
positrons.} of energy $E_e = 27.5$\,\gev\ and a proton beam of energy
$E_p =920$\,\gev.
 
A detailed description of the ZEUS detector{\ZcoosysfnCEeta} can be
found elsewhere~\cite{zeus:1993:bluebook}. Charged particles were
measured in the central tracking detector (CTD)~\citeCTD and, in
HERA-II, in a silicon microvertex detector~\cite{nim:a581:656}. These
operated in a magnetic field of $1.43$~T provided by a thin
superconducting solenoid.  The high-resolution uranium--scintillator
calorimeter (CAL)~\citeCAL consisted of three parts: the forward
(FCAL), the barrel (BCAL) and the rear (RCAL) calorimeters. The BCAL
covered the pseudorapidity range $-0.74$ to 1.10 as seen from the
nominal interaction point, and the FCAL and RCAL extended the coverage
to the range $-3.5$ to 4.0.  Each part of the CAL was subdivided into
elements referred to as cells. The barrel electromagnetic calorimeter
(BEMC) cells had a pointing geometry directed at the nominal
interaction point, and were approximately $5\times20\,\mathrm{cm^2}$
in cross section, with the finer granularity in the $Z$ direction and
the coarser in the $(X,Y)$ plane.  This fine granularity allows the
use of shower-shape distributions to distinguish isolated photons from
the products of neutral meson decays such as $\pi^0 \rightarrow
\gamma\gamma$.  The CAL energy resolution, as measured under test-beam
conditions, was $\sigma(E)/E = 0.18/\sqrt{E}$ for electrons and
$0.35/\sqrt{E}$ for hadrons, where $E$ is in\,\gev.

In most HERA events, the outgoing electron passes inside the inner
aperture of the RCAL, corresponding to a scattering angle of approximately
70 mrad with an upper limit on $Q^2$ of the order of 1 \gev$^2$.
The absence of a detected electron corresponds to a good approximation to
a photoproduction event.

During the HERA-I running, the aperture between the proton beam pipe
and the surrounding FCAL was occupied by the forward plug calorimeter
(FPC), which extended the rapidity coverage to +5.0~\cite{fpc}.  
In particular, it improved the reliability of the
measurement of the rapidity gap in diffractive events.  During the
HERA-II running, the configuration of this region was altered and a
beam-focusing magnet occupied the place of the FPC. 

The luminosity was measured \cite{lumi2} using the reaction $ep
\rightarrow e\gamma p$ by a luminosity detector which 
for HERA-I running consisted of a lead--scintillator calorimeter
\cite{desy-92-066,*zfp:c63:391,*acpp:b32:2025}
and for HERA-II running consisted of two independent systems: the
lead--scintillator calorimeter and a magnetic
spectrometer~\cite{nim:a565:572}.

\section{Effects of proton dissociation} 
\label{sec-dissoc}

Diffractive events are characterised by a rapidity gap between the
forward proton, or dissociated-proton system, and the rest of the
particles in the event.  A sample of diffractive events may be
obtained by excluding the events in which particles are recorded in the
forward regions of the detector beyond a maximum pseudorapidity value,
\etamax, taken as 2.5 in the present analysis.  The forward-scattered 
proton is not detected; however the accepted event sample includes contributions in
which the proton emerges in a dissociated state whose products pass
undetected inside the central aperture of the FCAL or FPC.  In some
cases, wider-angle dissociation products may be detected and cause the
event to fail the diffractive selections.

The HERA-I and HERA-II detector configurations differ in their ability
to identify events with proton dissociation.  For the HERA-I data, the
use of the FPC allowed most of these events to be rejected, but the
recorded cross sections still include a contribution from undetected
dissociated-proton systems with masses up to approximately 3\,\gev. In
the analysis of diffractive dijet events in photoproduction, this was
evaluated to be $16\pm4\%$ of the total published diffractive cross
section~\cite{epj:c55:177}.  For the HERA-II data, the size of the
central aperture of the FCAL was doubled. This, together with the
absence of the FPC and the possibility of secondary scattering from
the beam-focusing magnet, generated two effects which act in opposite
directions. In the first of these, the measured differential cross
sections include a larger contribution from proton dissociation; in
the analysis of diffractive dijet production in DIS, using
$\etamax=2.0,$ this contribution was evaluated to be $45\pm15\%$ and
comprises dissociated-proton systems with masses up to approximately
6\,\gev~\cite{epj:c76:1}.  A similar contribution would be expected in
diffractive photoproduction. It affects only the normalisation of the
differential cross sections, if the principle of vertex factorisation
is assumed to hold. However, it was possible for particles within a
dissociated-proton system to scatter from the focusing magnet into the
detector.  This effect was not accurately simulatable. It can reduce
the fraction of proton-dissociated events in the sample by removing
the forward rapidity gap in some of the events.

The higher statistics available in the HERA-II running made this data
set suitable for studying the distributions of kinematic variables,
which are described in Section~\ref{kin-q}.  However the possible
presence of a substantial number of events with proton dissociation
should be allowed for in the measured cross sections. The HERA-I data
set, being less affected by the proton dissociation and with the
focusing magnet absent, was used to evaluate an integrated ``visible''
cross section taken over the observed ranges of the measured
variables.

\section{Measured variables}
\label{kin-q}
All the measured quantities used in this analysis were determined in
the laboratory frame.  In direct photon processes in photoproduction,
the incoming virtual photon is absorbed by a quark from the target
particle, here a Pomeron, while in resolved photon processes, the
virtual photon's hadronic structure provides a quark or gluon that
interacts with a quark or gluon from the Pomeron.  These two classes
of process, which are unambiguously defined only at the leading order
(LO) of QCD, may be partially distinguished in events containing a
high-\ET\ photon and a jet by means of the quantity
\begin{equation}
\xgamm =  \frac{\Egam+\Ejet-\pzgam-\pzjet}
{E^{\text{all}}- p_Z^{\text{all}}}, 
\end{equation}
which measures the fraction of the incoming photon energy that is
given to the outgoing photon and jet.  The quantities \Egam\ and
\Ejet denote the energies of the outgoing photon and the jet,
respectively, and $p_Z$ denotes the corresponding longitudinal
momenta. The suffix ``all'' refers to all objects that are measured in
the detector or, in the case of simulations at the hadron level, all
final-state particles except for the scattered beam electron and the
outgoing proton. Events with a detected final-state electron are
excluded from this analysis.

 At LO, \xgamm= 1 for direct photon events, while resolved photon events can have 
any value in the range (0, 1). Direct photon events at higher
order can have \xgamm\ less than unity, but the presence of the LO
processes generates a prominent peak in the observed cross section at
high values of \xgamm.

When the  proton radiates a Pomeron that interacts
with an incoming photon, the fraction of the proton energy
carried by the radiated Pomeron is given to a good approximation by:
\begin{equation}
 \xpom =(E^{\text{all}}+ p_Z^{\text{all}})/2E_p,
\end{equation} 
where $E_p$ is the energy of the proton beam.  

The mass of the observed system, excluding the forward proton and its
possible dissociation products but including all the reaction products
of the incoming photon and Pomeron, is evaluated as:
\begin{equation}
M_{X} = \sqrt{(E^{\text{all}})^2 - (p_Z^{\text{all}})^2}. 
\end{equation}

The Pomeron may be described analogously to the
photon~\cite{zp:c65:657,mrw}. The fraction of the Pomeron energy that
takes part in the hard interaction that generates the outgoing photon
and jet is given by~\cite{epj:c70:15}:
\begin{equation} 
\zpom =  \frac{\Egam+\Ejet+\pzgam+\pzjet}{E^{\text{all}}+ p_Z^{\text{all}}},
\end{equation}
where the quantities are as before\footnote{The alternative formulation\;
$z_{\pom}^{\mathrm{obs}}= (\ETgam\exp{\etagam} +\ETjet\exp{\etajet})
/(E^{\text{all}}+p_Z^{\text{all}}),$\; where \ET\ denotes transverse
energy, yields equivalent results.}, and $\zpom=1$ corresponds to
direct Pomeron events, which are equivalent to the presence of a delta-function
in the PDFs at $\zpom=1$~\cite{mrw,zp:c65:657}.
An event whose observed final
state consists only of a prompt photon and a jet has $\xgamm=\zpom=1.$

Further variables that are used are as follows.  A measurable
approximation for the fraction $y$ of the incoming electron energy that is
transferred to the exchanged virtual photon is the
Jacquet--Blondel variable, $y_{\JB}$~\cite{yjb}, where in the present analysis
\begin{equation}
y_{\JB} = \sum\limits_i E_i(1-\cos\theta_i)/2E_e.
\end{equation}
Here,  $E_i$ is the energy
of the $i$-th CAL cell, $\theta_i$ is its polar angle and the sum runs
over all cells~\cite{pl:b303:183}. 
The photon--proton centre-of-mass energy, $W$, is calculated as 
\begin{equation}
   W = \sqrt{4yE_pE_e + m_p^2},
\end{equation}
where the small finite value of $Q^2$ is neglected, $m_p$ is the
proton mass and, at the detector level, $y$ is replaced by $y_{\JB}$.



\section{Monte Carlo event simulation} 
\label{sec-mc} 
Monte Carlo (MC) event samples were employed to model signal and
background processes.  The generated MC events were passed through the
ZEUS detector and trigger simulation programs based on {\sc Geant}
3~\cite{tech:cern-dd-ee-84-1}. They were then reconstructed and
ana\-lysed using the same programs as used for the data.  The effects
of the beam-focusing magnet in HERA-II were not well modelled in the
ZEUS apparatus simulation.

\subsection{\sc Rapgap}
The program \RAPGAP~3.2~\cite{rapgap,jung} was used to simulate the
diffractive process $ep \to ep\gamma X$, where $X$ denotes the
presence of final-state hadrons. In addition to enabling acceptance
corrections and event-reconstruction efficiencies to be calculated,
\RAPGAP\ also provided a physics model to compare to the results of
the present measurements. In \RAPGAP, the incoming photon is radiated
from the electron using the equivalent-photon approximation.  The
Pomeron carries a fraction \xpom\ of the proton longitudinal momentum
and is modelled as a hadron-like state within the framework of the
factorisation hypothesis of Ingelman and Schlein~\cite{ingelman}.  In
direct photon processes, it is assumed in \RAPGAP\ that the incoming
photon scatters elastically off a quark in the resolved Pomeron.  In
resolved photon processes, gluon--quark and antiquark--quark scattering
produce an outgoing photon and a jet.  Hadronisation of the outgoing
partons is performed using
\PYTHIA 6.410~\cite{jhep:0605:026}.

Event samples were generated for direct and resolved photon
interactions with a resolved Pomeron.  The default parameters were
used and the $\alpha_s$ scale was $p_{T\,\gamma}^2$, where
$p_{T\,\gamma}$ is the transverse momentum of the outgoing photon.
The selected PDF sets were, for the Pomeron, H1 2006 DPDF Fit
B~\cite{epjc:48:715} and, for the resolved photon,
SASGAM-2D~\cite{plb:376:193}.  Proton dissociation was not generated
in the present analysis.  In the original QCD fit by H1 using DIS
data~\cite{epjc:48:715}, resolved Pomeron PDFs were obtained for
$\zpom< 0.8$;
\RAPGAP\ uses these with an extrapolation to cover the entire
\zpom\ range up to 1.0. Since a simulation of the type of process in Fig.~1(c) 
was not available, the simulation by \RAPGAP\ was used throughout.

\subsection{Background simulations}
A background to the isolated photons measured here comes from neutral
mesons in hadronic jets, in particular $\pi^0$ and $\eta$, where meson
decay products can create an energy cluster in the BEMC that passes
the experimental selection criteria for a photon candidate. To model these
effects, \RAPGAP\ was used to generate direct and resolved
diffractive scattering that produced exclusive two-jet events that did
not contain prompt photons in the final state.  These were analysed
using the same program chain as for the prompt photon events.

A separate potential source of background came from non-diffractive
prompt photon events; \PYTHIA\-6.416 was used to generate 
processes of this type,  making use of the CTEQ4~\cite{pr:d55:1280} and
GRV~\cite{grv} proton and photon PDF sets.

For additional background studies, Bethe--Heitler (BH) event samples
were obtained using the \GRAPE\ MC~\cite{grape}.  DIS event samples
with initial-state photon radiation were also generated using the \GRAPE\
and the \DJANGOH~6 programs~\cite{djangoh} interfaced with
\ARIADNE~\cite{ariadne}.


\section{Event selection}
\label{sec-selec}

The basic event selection and reconstruction was performed as
previously~\cite{pl:b730:293}.  A three-level trigger system was used
to select events online
\cite{zeus:1993:bluebook,uproc:chep:1992:222,nim:a580:1257}:
\begin{itemize}
\item the first-level trigger required a loosely measured track in the CTD
and energy deposited in the CAL that included conditions to select an
isolated electromagnetic signal;
\item at the second level, conditions for an event with at least 8\,\gev\ 
of summed transverse energy were imposed;
\item at the third level, the event was reconstructed and a high-energy 
photon candidate was required.
\end{itemize}

In the offline event analysis, some general conditions were applied as
follows:
\begin{itemize}
\item to reduce background from non-$ep$
collisions, events were required to have a reconstructed vertex
position, $Z_{\mathrm{vtx}}$, within the range $|Z_{\mathrm{vtx}}|<
40\,\mathrm{cm}$;  
\item no identified electron with energy above 3.5\,\gev\ was allowed in the 
event;
\item at least one vertex-fitted track with $p_T > 0.2$\,\gev\ was required;
\item the accepted range of incoming virtual photon energies was defined 
by the requirement $0.2 < y_{\JB} < 0.7$.  The lower cut strengthened the
trigger requirements and the upper cut suppressed DIS events;
\item a potential source of unwanted events arises from BH 
processes of the type $ep\to ep\gamma$, where the outgoing electron is
at a wide angle and any possible dissociation
products of the proton are not observed in the detector.  If collinear
initial-state radiation from the beam electron takes place, these
events may be recorded within the allowed $y_{\JB}$ range, the
outgoing electron being interpreted as a jet. Such events have a small
number of outgoing particles in the detector, and are efficiently
rejected by the veto on identified electrons, and by a further
requirement that the number of energy-flow objects in the event (see
below) with energy above 0.2\,\gev\ must exceed 5.  The rejection
efficiency for these events was close to 100\% and was verified by
means of event samples from \GRAPE\ that simulated the
BH processes. The kinematically similar deeply virtual Compton
scattering (DVCS) process was excluded in the same way as the BH
processes.  Approximately 2\% of \RAPGAP\ events were rejected by this
selection. The procedure was further checked by a visual scan of the
data events.
\end{itemize}

The event analysis made use of energy-flow objects
(EFOs)~\cite{epj:c1:81,*epj:c6:43}, which were constructed from
clusters of calorimeter cells, associated with tracks when
appropriate. Tracks not associated with calorimeter clusters were also
included. EFOs with no associated track, and with at least $90\%$ of
the reconstructed energy measured in the BEMC, were taken as photon
candidates. Photon candidates with wider electromagnetic showers than
are typical for a single photon were accepted at this stage so as to
make possible the evaluation of backgrounds.  The photon energy scale
was calibrated~\cite{pl:b730:293,andrii} by means of an analysis of DVCS
events recorded by ZEUS, in which the detected final-state particles
comprised a scattered electron, whose energy measurement is well
understood, and a balancing outgoing photon.

Jet reconstruction was performed making use of all the EFOs in the
event, including photon candidates, by means of the $k_T$ clustering
algorithm~\cite{np:b406:187} using the $E$-scheme in the
longitudinally invariant inclusive mode~\cite{pr:d48:3160} with the
radius parameter set to 1.0.  One of the jets found by this procedure
corresponds to or includes the photon candidate. An accompanying jet
was used in the analysis; if more than one jet was found, that with
the highest transverse energy, $E_T^{\mathrm{jet}},$ was selected.  In the
kinematic region used, the resolution of the jet transverse energy was
about 15--20\%, estimated using MC simulations.

To reduce the contribution of photons from fragmentation processes,
and also the background from the decay of neutral mesons within jets,
the photon candidate was required to be isolated from other hadronic
activity.  This was imposed by requiring that the photon-candidate EFO
had at least 90\% of the total energy of the reconstructed jet of
which it formed a part, a condition that was imposed also in the
hadron-level calculations. High-\ET\ photons radiated from scattered
leptons were further suppressed by rejecting photons that had a nearby
track. This was achieved by demanding $\Delta R > 0.2,$ where 
\begin{equation}
\Delta R = \sqrt{(\Delta \phi)^2 + (\Delta\eta)^2}
\end{equation}
and  is the distance to the
nearest reconstructed track with momentum greater than
$250\,\mathrm{MeV}$ in the $\eta \-- \phi$ plane, where $\phi$ is the
azimuthal angle. This latter condition was applied at the
detector level for both MC and for data.

The final event selection was as follows:
\begin{itemize}
\item each event was required to contain a photon candidate with a 
reconstructed transverse energy, $E_T^{\gamma}$, in the range $5
<E_T^{\gamma}<15\,$\gev\ and with pseudorapidity,
$\eta^{\gamma}$, in the range $-0.7 < \eta^{\gamma} < 0.9$;
\item  a hadronic jet was required to have  
$E_T^{\mathrm{jet}}$ between 4 and 35\,\gev\ and to lie within the
pseudorapidity, $\eta^{\mathrm{jet}}$, range $-1.5
<\eta^{\mathrm{jet}} < 1.8$;
\item 
the maximum pseudorapidity for EFOs with energy above 0.4\,\gev,
\etamax, was required to satisfy $\eta_{\mathrm{max}} < 2.5$ in order
to select diffractive events, characterised by a large rapidity gap;

\item 
a requirement $\xpom < 0.03$ was made to reduce further any
contamination from non-diffractive events;

\item 
the energy deposited in the FPC was required to be less than 1\gev\ for
the HERA-I data sample~\cite{kagawa}.

\end{itemize} 

\section{Extraction of the photon signal}

The selected samples contain a substantial admixture of background
events in which one or more neutral mesons, such as $\pi^0$ and
$\eta$, have decayed to photons, thereby producing a photon candidate
in the BEMC.  The photon signal was extracted statistically following
the approach used in previous ZEUS analyses
\cite{pl:b595:86,pl:b687:16,pl:b715:88,pl:b730:293}.  
The method made use of the
energy-weighted width, measured in the $Z$ direction, of the BEMC
energy cluster comprising the photon candidate. This width was calculated
as 
\begin{equation}
\langle\delta Z\rangle=
\sum \limits_i E_i|Z_i-Z_{\mathrm{cluster}}|
\mathrm{\,\left/\,\right.}( w_{\mathrm{cell}}\sum \limits_i E_i),
\end{equation}
 where $Z_{i}$ is
the $Z$ position of the centre of the $i$-th cell,
$Z_{\mathrm{cluster}}$ is the energy-weighted centroid of the EFO
cluster, $w_{\mathrm{cell}}$ is the width of the cell in the $Z$
direction, and $E_i$ is the energy recorded in the cell. The sum runs
over all BEMC cells in the EFO cluster.  

\begin{figure}[p]
\begin{center}
\includegraphics[width=0.7\linewidth]{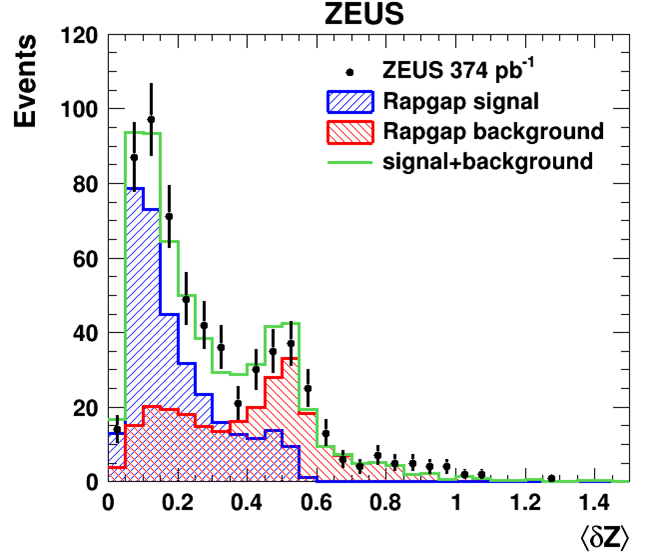}
\end{center}
\caption{\small 
Distribution of $\langle \delta Z \rangle$ for selected diffractive
events with a photon candidate and at least one jet, for the full sample of HERA-II
data. The error bars denote the statistical uncertainties on the data,
which are compared to the fitted signal and background components from
the MC.  The unit of measurement of $\langle \delta Z \rangle$ is the
width of one BEMC cell.  }
\label{fig:showers}
\end{figure}

The number of isolated-photon events in the data was determined by a
binned maximum-likelihood fit to the $\langle\delta Z \rangle$
distribution in the range $0.05<\langle \delta Z \rangle < 0.8$,
varying the relative fractions of the signal and background components
as represented by histogram templates obtained from the MC.  The fit
was performed for each measured cross-section interval, with $\chi^2$
values of typically 1.0 per degree of freedom.
Figure~\ref{fig:showers} shows the fitted $\langle\delta Z\rangle$
distribution for the full sample of selected HERA-II events with a
photon candidate and at least one jet.  The peak seen at $\langle
\delta Z \rangle \sim 0.5$ is due to $\pi^0$ decays.

For the HERA-I data
sample, starting from 161 (127) selected events containing a photon
candidate without (with) at least one accompanying jet, the fit gave 91 (76)
photon events.  For the HERA-II data sample, the respective figures
were 767 (598) selected events, giving 366$\pm$31 (311$\pm$28) photon
events after the fit.  It is apparent that a large fraction of the
isolated hard photons are accompanied by one or more observed jets.

\section{Event distributions and evaluation of cross sections}

\begin{figure}
\begin{center}
\includegraphics[width=0.7\linewidth]{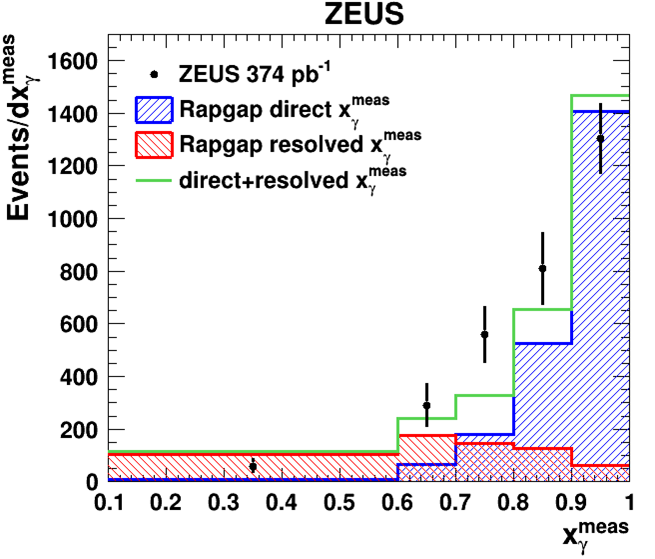}  
\end{center}
\caption{\small  
HERA-II events with a photon and at least one jet as a function of \xgamm, per
unit interval in \xgamm, compared to a normalised 70:30 mixture of
direct:resolved photon \RAPGAP\ events without reweighting.\protect\\[10mm]}
\label{fig:xgam}
\end{figure}

After applying the selections described above, event distributions
were extracted for the HERA-II data.  The distribution of events in
\xgamm\ is shown in Fig.~\ref{fig:xgam}.  A
70:30 mixture of direct:resolved photon events generated with \RAPGAP\
gives a reasonable description of the data and was employed in the
following analysis.  This applies both for the full data set and for
the two separate ranges of \zpom\ that are described below.

Figure~\ref{fig:rew}(a) shows the event distribution in \zpom,
together with the prediction obtained from \RAPGAP. 
\RAPGAP\ describes the distribution well for
$\zpom < 0.9$, but above this value the data lie above the
\RAPGAP\ prediction. Here, \RAPGAP\ does not simulate all applicable 
physics processes, such as the type illustrated in Fig.~1(c).  A good
description of the data is required in order to calculate acceptances;
in order to obtain this, a weighting factor of 7.0 may be applied to
the direct photon component of \RAPGAP\ for hadron-level values of
\zpom\ above 0.9. The observed \zpom\ distribution is then well 
described, and Fig.~\ref{fig:rew}(b) shows
that the reweighted \RAPGAP\ also provides better agreement with the
\etamax\ event distribution (see Sections 
\ref{sec-dissoc} and \ref{sec-selec}).  For the other measured variables,
the two \RAPGAP\ descriptions are both good and are generally similar,
with no clear discrimination between them.  The experimental cross
sections were evaluated using acceptances that used the reweighted
version of \RAPGAP as described above.

A bin-by-bin correction method was used to determine the differential 
cross section in a given variable, by means of the equation
\begin{equation}
\frac{d\sigma}{dY} = \frac{\mathcal{A}\, N(\gamma)}{  
\mathcal{L} \, \Delta Y},
\end{equation}
where $N(\gamma)$ is the number of photons in a bin as extracted from
the $\langle\delta Z\rangle$ fit, $\Delta Y$ is the bin width,
$\mathcal{L}$ is the total integrated luminosity, and $\mathcal{A}$ is
a correction given by the reciprocal of the acceptance.  The
correction $\mathcal{A}$ was calculated, using \RAPGAP\ samples, as
the ratio of the number of events that were generated in the given
bin, according to the chosen definitions, divided by the number of
events obtained in the bin after event reconstruction and selection as
for the data. As a check on the bin-by-bin correction method, an
expectation-maximisation unfolding technique~\cite{unfold} was applied
and gave similar results.

After the background subtraction, it was found that of
the events with a photon and at least one jet, approximately 5\% 
of those with $\zpom<0.9$ had a second accepted jet.  The number 
of events with a third accepted jet was consistent with zero. 
No additional jets are expected in events with $\zpom\ge0.9,$
owing to kinematic constraints, and none were found.

\section{Systematic uncertainties}
\label{sec:syst}

The main sources of systematic uncertainty on the measured visible HERA-II
cross sections were evaluated as follows:
\begin{itemize}
\item 
the energy of the photon candidate was varied by $\pm 2\%$ in the MC
at the detector level, and independently the energy of the
accompanying jet was varied by $\pm2\%$.  These variations represent
the energy scale uncertainties~\cite{andrii}.  Each of them gave
variations in the measured cross sections of typically $\pm5\%$;
\item 
the uncertainty in the acceptance due to the estimation of the
relative fractions of direct photon and resolved photon events in the
\RAPGAP\ MC sample was estimated by varying the fraction of direct
photon events between 60\% and 80\%; the changes in the cross sections
were typically $\pm2\%$;
\item  
the dependence of the result on the modelling by the MC of the
hadronic background in the $\langle\delta Z\rangle$ distribution was
investigated by varying the upper limit for the $\langle
\delta Z\rangle$ fit in the range $[0.6, 1.0]$~\cite{pl:b715:88}; this gave 
a $\pm  2\%$ variation;
\item  
the non-diffractive photoproduction background was estimated by
fitting a number of experimental variables to mixtures of \RAPGAP\ and
\PYTHIA\ samples.  The \PYTHIA\ samples were treated in the same way
as the data, using an appropriate mixture of resolved and direct
photoproduction events. It was found that a satisfactory description
of the data was obtained with no non-diffractive background, but that
up to 10\% of background could not be excluded, as illustrated in
Fig.~\ref{fig:rew}. This is included as an asymmetric systematic
uncertainty.
\end{itemize}
The uncertainties listed above were combined in quadrature.
The normalisation issues due to proton dissociation as discussed in
Section 3 were not further evaluated as they do not affect the shape of
the distributions.
 
A possible contamination of DIS events was
investigated using the programs \GRAPE\ and \DJANGOH,
and a possible contribution arising from photon--photon interactions
was investigated using \GRAPE. Both of these were found
to be negligible.  Other sources of systematic uncertainty that were
estimated to be negligible included the modelling of the
track-isolation cut and the track-momentum cut, and also the cuts on
photon isolation, the electromagnetic fraction of the photon shower,
$y_{\JB}$, and $Z_{\text{vtx}}$.

The uncertainties of 2.0\% on the trigger efficiency and 1.9\% on the
luminosity measurement were not included in the figures.  These
contributions are included in the uncertainties on the visible cross
sections determined from the HERA-I data, together with the other
systematic uncertainties evaluated as for the HERA-II cross sections.

\section{Results}
\label{sec:results}

Cross sections were measured for the diffractive production of an
isolated photon, inclusive and with at least one accompanying jet, in
the kinematic region defined by $Q^2< 1\,\mathrm{\gev}^2$, $0.2 < y <
0.7,$ $-0.7<\etagam < 0.9$, $5 < \ETgam< 15\,\mathrm{\gev} $, $4 <
\ETjet<35$\gev\ and $-1.5 <\etajet< 1.8$. The diffractive condition
required that $\etamax<2.5$ and $\xpom <0.03$.  As a result of the
removal of the BH and DVCS events, the measurements are sensitive only
to events with more than five observed final-state particles,
including the isolated photon.  This condition was imposed on the MC
events at the detector level but not at the hadron level.  All cross
sections were evaluated at the hadron level in the laboratory frame,
and the jets were formed according to the $k_T$ clustering algorithm
with the radius parameter set to 1.0. Both at the detector and hadron
levels, photon isolation was imposed by requiring that the photon
candidate had at least 90\% of the total energy of the reconstructed
jet of which it formed a part.  If more than one accompanying jet was
found within the designated \etajet range in an event, that with
highest \ETjet was taken. No subtraction for dissociated-proton states
has been made. As explained in Section~\ref{sec-dissoc},
these are uncertain and could amount to 40\% of the visible cross section.

With the above selections, the effect of the \etamax requirement is to
remove 64\% of the diffractive events with $\xpom <0.03$, as evaluated
using the \RAPGAP\ model. In order to avoid the large extrapolation
that would be needed to include the full \etamax\ range, and given the
additional presence in this range of larger non-diffractive
backgrounds and uncertain effects of proton dissociation, ``visible''
cross sections ­are quoted here for the range defined by $\etamax<2.5$
and $\xpom<0.03$.

Differential cross sections for inclusive prompt-photon production,
using the HERA-II data, are shown for the quantities \ETgam\ and
\etagam\ in Fig.\ \ref{fig:xsincl}(a, b). Differential cross sections
for the quantities \xpom\ and \Mx\ for events with an inclusive prompt
photon are shown in Fig.~\ref{fig:xsincl}(c, d). The predictions of
\RAPGAP, normalised to the data, are in good agreement with the data
in both the unreweighted and reweighted cases. The data are listed 
in Tables \ref{tab:etgi}--\ref{tab:mxi}.

For events containing a photon and at least one jet, the differential
cross section as a function of \zpom\ is plotted in
Fig.~\ref{fig:xszpom} and listed in Table \ref{tab:zpom}. It shows
evidence for an excess of data above the nominal prediction of
\RAPGAP\ for $\zpom \geq 0.9$, which lies beyond the region where the
Pomeron PDFs were originally evaluated. As a check on this result, the
analysis was repeated with the selection on \etamax\ removed and
applying different selections on
\xpom. These variations had the effect of changing the measured shape
of the cross section $d\sigma/d\zpom$ for $\zpom<0.9$, but the excess
above the \RAPGAP\ prediction for $\zpom\geq0.9$ remained present in each
case.

Figures~\ref{fig:xsgam}--\ref{fig:xsphe}, together with Tables
\ref{tab:etgj}--\ref{tab:etamax}, show the differential cross sections 
for a number of kinematic variables for the full \zpom\ range and
separately for the ranges $\zpom < 0.9$ and $\zpom\geq0.9$. The
variables presented are the transverse energy and the pseudorapidity
of the photon and the jet, the incoming photon--proton centre-of-mass
energy, $W,$ the ratio of the transverse energies of the photon and
the jet, the quantities \xgamm, \xpom\ and \Mx, the differences in
azimuth and pseudorapidity of the photon and the jet, $\delphi =
|\phi^\gamma -\phi^{\mathrm jet}|$ and $\Deleta = \eta^\gamma -
\etajet$, and \etamax.  Cross sections for $\ETjet$ above 15\gev\ are
omitted from Figs.~\ref{fig:xsjet}(a)--(c) owing to limited
statistics, but the data in this range are included in the other
cross-section measurements.

The distributions shown in Figs.~\ref{fig:xsgam}--\ref{fig:xsphe} are
generally well described by \RAPGAP, apart from \etamax\ in
Fig.~\ref{fig:xsphe}(g) when \RAPGAP\ is not reweighted.  For
$\zpom<0.9,$ \RAPGAP\ normalised to the data in this range is in
agreement with the data in all variables. For $\zpom\geq0.9,$ \RAPGAP\
gives a good phenomenological description of the shape of the
data. The distribution in \delphi\ confirms that the data are
dominated by events with a photon and one jet. This is also confirmed
by the distribution of the ratio of the transverse energies of the
photon and the jet.

The cross-section distribution in \zpom\ may be compared to the
results obtained by ZEUS for the diffractive production of dijet
systems~\cite{epj:c55:177,kagawa}, where the photoproduction data 
are not well described by \RAPGAP\ but do 
not show a similar rise at high values of \zpom.
In DIS, the diffractive production of
exclusive dijets was found to be better described by a
two-gluon-exchange, or direct Pomeron, model than by
\RAPGAP~\cite{epj:c76:1}.  The present prompt-photon results give
evidence for the presence of a direct Pomeron process in
diffractive photoproduction with $\zpom\geq0.9$. Events in
this region show indications of a resolved photon contribution
(Fig.~\ref{fig:xsxm}(c)), but are dominated by direct photon
interactions.
This is the first measurement in this channel. At present, no theoretical model is 
available that might give a quantitative prediction for this effect.

The integrated visible HERA-II cross sections for the diffractive
production of a prompt photon in the above kinematic region,
inclusively and with at least one jet, are found to be
$1.21\pm0.10^{+0.10}_{-0.16}$ pb and $1.14\pm0.10^{+0.07}_{-0.15}$ pb,
respectively. The smaller and calculable proton
dissociation contribution in the HERA-I data allows a correction for
this effect to be made.  Using the HERA-I data, analysed as for the
present HERA-II measurements, integrated cross sections of
$1.21\pm0.19^{+0.14}_{-0.14}$ pb and $1.10\pm0.19^{+0.09}_{-0.13}$ pb,
respectively, were obtained. These were evaluated with the same event
selections and kinematic limits as for the HERA-II measurements but
supplemented by a veto on events with an FPC signal of more than
1\gev.

The integrated cross section from the HERA-I data, evaluated with the
present experimental selections in the range $\zpom < 0.9$, is found
to be $0.68\pm0.14^{+0.06}_{-0.07}$ pb, with no allowance for proton
dissociation.  This becomes $0.57\pm0.12^{+0.05}_{-0.06}$ pb after
multiplying by a dissociation correction factor of 0.84. The
corresponding value from \RAPGAP\ is $0.68$~pb, with no proton
dissociation and no resolved-suppression
factor\,\cite{Collins,epj:c21:521}.  The agreement in shape and
normalisation found with the \RAPGAP\ predictions in the lower \zpom\
range, obtained using Pomeron PDFs generated from DIS data, is
consistent with a common set of Pomeron PDFs in the photoproduction
and DIS regimes.

\section{Conclusions}

The diffractive photoproduction of isolated photons, with and without
at least one accompanying jet, has been measured for the first time
with the ZEUS detector at HERA, using integrated luminosities of
$82\,\pm2\,\mathrm{pb}^{-1}$ from HERA-I and $374\pm7$ pb$^{-1}$ from
HERA-II.  Cross sections are presented in a kinematic region defined
in the laboratory frame by:\mbox{ $Q^2< 1 \,$} $\mathrm{\gev}^2$, $0.2
< y < 0.7$, $-0.7<\eta^\gamma < 0.9$, $5 < E_T^{\gamma}<
15\,\mathrm{\gev}$, $4< \ETjet <35$\,\gev\ and $-1.5 <\etajet<1.8$.
The diffractive requirement was $\etamax < 2.5$ and $\xpom< 0.03$.
Photon isolation was imposed by requiring that the photon had at least
$90\%$ of the energy of the reconstructed jet of which it formed a
part.

Differential cross sections are presented in terms of the transverse
energy and pseudorapidity of the prompt photon and the jet, and for a
number of variables that describe the kinematic properties of the
diffractively produced system, in particular the fraction of the
Pomeron energy given to the prompt photon and the jet, \zpom.
%
%

The data are compared with a standard \RAPGAP\ model that simulates
direct and resolved photon interactions with a resolved Pomeron.  With
the exception of \etamax\ and \zpom, the distributions in all the
variables are well described in shape by this model over the whole
\zpom\ range and in the ranges $\zpom<0.9$ and $\zpom\geq0.9$
separately.  For $\zpom\geq0.9$, there is evidence for an excess in
the data above the nominal \RAPGAP\ prediction.  This excess indicates
the presence of a direct Pomeron interaction, and is observed
predominantly in the direct photon channel.

\section*{Acknowledgements} \label{sec-ack} \Zacknowledge\ We also
thank H. Jung for providing invaluable assistance with the use of
\RAPGAP.  \vfill\eject 

\raggedright{
\providecommand{\etal}{et al.\xspace}
\providecommand{\coll}{Collaboration}
\catcode`\@=11
\def\@bibitem#1{%
\ifmc@bstsupport
  \mc@iftail{#1}%
    {;\newline\ignorespaces}%
    {\ifmc@first\else.\fi\orig@bibitem{#1}}
  \mc@firstfalse
\else
  \mc@iftail{#1}%
    {\ignorespaces}%
    {\orig@bibitem{#1}}%
\fi}%
\catcode`\@=12

}
\begin{table}
\begin{center}
\begin{tabular}{|rcr|c|}
\hline
\multicolumn{3}{|c|}{$E_T^{\gamma}$ range}   &       \\[-1.8ex]
\multicolumn{3}{|c|}{ (GeV)}   &  
 \raisebox{1.8ex}{$\frac{d\sigma}{dE^{\gamma}_T}$ ($\mathrm{pb}\,\mathrm{GeV}^{-1})$} \\    
\hline
5.0 &--& 6.0   & $0.549\pm 0.087\,\mathrm{(stat.)}\,^{+0.033}_{-0.072}\,\mathrm{(syst.)}$ \\
6.0 &--& 7.0   & $0.269\pm 0.054\,\mathrm{(stat.)}\,^{+0.038}_{-0.038}\,\mathrm{(syst.)}$ \\
7.0 &--& 8.0   & $0.187\pm  0.032\,\mathrm{(stat.)}\,^{+0.023}_{-0.027}\,\mathrm{(syst.)}$ \\
8.0 &--& 15.0   & $0.031\pm0.005\,\mathrm{(stat.)}\,^{+0.004}_{-0.005}\,\mathrm{(syst.)}$ \\
\hline

\end{tabular}
\end{center}
\caption{Differential cross-section $\frac{d\sigma}{dE^{\gamma}_T}$ 
for inclusive photons in diffractive photoproduction.  (Figure~\ref{fig:xsincl}(a))
}
\label{tab:etgi}
\end{table}
\begin{table}
\begin{center}
\begin{tabular}{|rcr|c|}
\hline
\multicolumn{3}{|c|}{ $\eta^{\gamma}$ range } 
          &  $\frac{d\sigma}{d\eta^{\gamma}}$ ($\mathrm{pb}$) \\[0.5mm]
\hline
--\,0.7 &--& --\,0.3   & $1.33 \pm  0.19\,\mathrm{(stat.)}\,^{+0.13}_{-0.18}\,\mathrm{(syst.)}$ \\
--\,0.3 &--& 0.1   & $0.87 \pm  0.14\,\mathrm{(stat.)}\,^{+0.09}_{-0.11}\,\mathrm{(syst.)}$ \\
0.1 &--& 0.5   & $0.419 \pm  0.105\,\mathrm{(stat.)}\,^{+0.019}_{-0.057}\,\mathrm{(syst.)}$ \\
0.5 &--& 0.9   & $0.485 \pm  0.095\,\mathrm{(stat.)}\,^{+0.056}_{-0.064}\,\mathrm{(syst.)}$ \\
\hline
\end{tabular}
\end{center}
\caption{Differential cross-section $\frac{d\sigma}{d\eta^{\gamma}}$ 
for inclusive photons in diffractive photoproduction. (Figure~\ref{fig:xsincl}(b))
}
\label{tab:etagi}
\end{table}

\begin{table}
\begin{center}
\begin{tabular}{|rcr|c|}
\hline
\multicolumn{3}{|c|}{ $\xpom$ range }  &  $\frac{d\sigma}{d \xpom}$ ($\mathrm{pb}$) \\[0.5mm]
\hline
0.0 &--&  0.005   & $24.3 \pm  7.5\,\mathrm{(stat.)}\,^{+2.0}_{-3.9}\,\mathrm{(syst.)}$ \\
0.005 &--& 0.01   & $87.6 \pm  12.8\,\mathrm{(stat.)}\,^{+7.8}_{-11.1}\,\mathrm{(syst.)}$ \\
0.01 &--& 0.015   & $67.1 \pm  10.9\,\mathrm{(stat.)}\,^{+5.5}_{-11.0}\,\mathrm{(syst.)}$ \\
0.015 &--&  0.02   & $41.9 \pm  8.6\,\mathrm{(stat.)}\,^{+3.3}_{-5.8}\,\mathrm{(syst.)}$ \\
0.02 &--& 0.025   & $15.4 \pm  4.7\,\mathrm{(stat.)}\,^{+1.3}_{-2.0}\,\mathrm{(syst.)}$ \\
0.025 &--& 0.03   & $4.9 \pm  3.2\,\mathrm{(stat.)}\,^{+1.5}_{-0.9}\,\mathrm{(syst.)}$ \\
\hline
\end{tabular}
\end{center}
\caption{Differential cross-section $\frac{d\sigma}{d \xpom}$ 
for inclusive photons in diffractive photoproduction.(Figure~\ref{fig:xsincl}(c))
}
\label{tab:xpi}
\end{table}

\begin{table}
\begin{center}
\begin{tabular}{|rcr|c|}
\hline
\multicolumn{3}{|c|}{$M_X$ range}   &       \\[-1.8ex]
\multicolumn{3}{|c|}{ (GeV)}   &  
 \raisebox{1.8ex}{$\frac{d\sigma}{dM_X}$ ($\mathrm{pb}\,\mathrm{GeV}^{-1})$} \\    
\hline

10.0 &--& 15.0   & $0.048\pm 0.008\,\mathrm{(stat.)}\,^{+0.006}_{-0.006}\,\mathrm{(syst.)}$ \\
15.0 &--& 20.0   & $0.101\pm 0.014\,\mathrm{(stat.)}\,^{+0.010}_{-0.015}\,\mathrm{(syst.)}$ \\
20.0 &--& 25.0   & $0.053\pm  0.009\,\mathrm{(stat.)}\,^{+0.010}_{-0.009}\,\mathrm{(syst.)}$ \\
25.0 &--& 30.0   & $0.029\pm0.007\,\mathrm{(stat.)}\,^{+0.002}_{-0.005}\,\mathrm{(syst.)}$ \\
30.0 &--& 40.0   & $0.005\pm0.002\,\mathrm{(stat.)}\,^{+0.001}_{-0.001}\,\mathrm{(syst.)}$ \\
\hline

\end{tabular}
\end{center}
\caption{Differential cross-section $\frac{d\sigma}{dM_X}$ 
for inclusive photons in diffractive photoproduction. (Figure~\ref{fig:xsincl}(d))
}
\label{tab:mxi}
\end{table}


\begin{table}
\begin{center}
\begin{tabular}{|rcr|c|}
\hline
\multicolumn{3}{|c|}{ $\zpom$ range }  &  $\frac{d\sigma}{d\zpom}$ ($\mathrm{pb}$) \\[0.5mm]
\hline
0.0 &--&  0.4      & $0.25 \pm 0.08\,\mathrm{(stat.)}\,^{+0.01}_{-0.04}\,\mathrm{(syst.)}$ \\
0.4   &--& 0.5    & $0.74 \pm 0.29\,\mathrm{(stat.)}\,^{+0.10}_{-0.15}\,\mathrm{(syst.)}$ \\
0.5  &--& 0.6     & $1.12 \pm 0.32\,\mathrm{(stat.)}\,^{+0.05}_{-0.14}\,\mathrm{(syst.)}$ \\
0.6   &--& 0.7    & $1.73 \pm 0.35\,\mathrm{(stat.)}\,^{+0.12}_{-0.24}\,\mathrm{(syst.)}$ \\
0.7  &--& 0.8     & $1.44 \pm 0.29\,\mathrm{(stat.)}\,^{+0.15}_{-0.15}\,\mathrm{(syst.)}$ \\
0.8   &--& 0.9    & $1.02 \pm  0.27\,\mathrm{(stat.)}\,^{+0.13}_{-0.19}\,\mathrm{(syst.)}$ \\
0.9   &--& 1.0    & $4.79 \pm  0.65\,\mathrm{(stat.)}\,^{+0.83}_{-0.93}\,\mathrm{(syst.)}$ \\
\hline
\end{tabular}
\end{center}
\caption{Differential cross-section $\frac{d\sigma}{d\zpom}$ 
for photons accompanied by at least one jet in diffractive photoproduction. (Figure~\ref{fig:xszpom})
}
\label{tab:zpom}
\end{table}

\begin{table}
\begin{center}
\begin{tabular}{|rcr|c|}
\hline
\multicolumn{3}{|c|}{$E_T^{\gamma}$ range}   &       \\[-1.8ex]
\multicolumn{3}{|c|}{ (GeV)}   &  \raisebox{1.8ex}{$\frac{d\sigma}{dE^{\gamma}_T}$ ($\mathrm{pb}\,\mathrm{GeV}^{-1})$}   \\ 
\hline\hline
\multicolumn{3}{|l|}{$0<\zpom\leq 1.0$}   & \\
\hline

5.0 & -- & 6.0   & $0.483 \pm  0.081\,\mathrm{(stat.)}\,^{+0.021}_{-0.067}\,\mathrm{(syst.)}$ \\
6.0 & -- & 7.0   & $0.257\pm  0.052\,\mathrm{(stat.)}\,^{+0.024}_{-0.030}\,\mathrm{(syst.)}$ \\
7.0 & -- & 8.0   & $0.185 \pm  0.033\,\mathrm{(stat.)}\,^{+0.025}_{-0.026}\,\mathrm{(syst.)}$ \\
8.0 & -- & 15.0  & $0.031 \pm  0.005\,\mathrm{(stat.)}\,^{+0.004}_{-0.004}\,\mathrm{(syst.)}$ \\
\hline
\multicolumn{3}{|l|}{$\zpom<0.9$}   & \\
\hline
5.0 & -- & 6.0   & $0.314 \pm  0.052\,\mathrm{(stat.)}\,^{+0.009}_{-0.040}\,\mathrm{(syst.)}$ \\
6.0 & -- & 7.0   & $0.143 \pm  0.034\,\mathrm{(stat.)}\,^{+0.015}_{-0.016}\,\mathrm{(syst.)}$ \\
7.0 & -- & 8.0   & $0.122 \pm  0.026\,\mathrm{(stat.)}\,^{+0.012}_{-0.015}\,\mathrm{(syst.)}$ \\
8.0 & -- & 15.0  & $0.014 \pm  0.003\,\mathrm{(stat.)}\,^{+0.001}_{-0.002}\,\mathrm{(syst.)}$ \\
\hline
\multicolumn{3}{|l|}{$\zpom\geq0.9$}   & \\
\hline
5.0 & -- & 6.0   & $0.112 \pm  0.044\,\mathrm{(stat.)}\,^{+0.023}_{-0.029}\,\mathrm{(syst.)}$ \\
6.0 & -- & 7.0   & $0.118 \pm  0.035\,\mathrm{(stat.)}\,^{+0.025}_{-0.023}\,\mathrm{(syst.)}$ \\
7.0 & -- & 8.0   & $0.056 \pm  0.018\,\mathrm{(stat.)}\,^{+0.017}_{-0.014}\,\mathrm{(syst.)}$ \\
8.0 & -- & 15.0  & $0.015 \pm  0.003\,\mathrm{(stat.)}\,^{+0.003}_{-0.002}\,\mathrm{(syst.)}$ \\
\hline
\end{tabular}
\end{center}
\caption{Differential cross-section
$\frac{d\sigma}{dE^{\gamma}_T}$ for photons accompanied by at least one jet
in diffractive photoproduction.  Here and below, the differences between 
the results evaluated for the entire \zpom\ range and the sum of the
corresponding results for the two partial ranges are of statistical origin.
(Figure~\ref{fig:xsgam}(a--c))
}
\label{tab:etgj}
\end{table}
\begin{table}
\begin{center}
\begin{tabular}{|rcr|c|}
\hline
\multicolumn{3}{|c|}{ $\eta^{\gamma}$ range } &\multicolumn{1}{|c|}{ $\frac{d\sigma}{d\eta^{\gamma}}$ 
($\mathrm{pb}$)}     \\[0.5mm]
\hline\hline
\multicolumn{3}{|l|}{$0<\zpom\leq 1.0$}   & \\
\hline
--0.7 & -- & --\,0.3   & $1.24 \pm  0.18\,\mathrm{(stat.)}\,^{+0.08}_{-0.16}\,\mathrm{(syst.)}$ \\
--\,0.3 & -- & 0.1   & $0.78 \pm  0.13\,\mathrm{(stat.)}\,^{+0.06}_{-0.10}\,\mathrm{(syst.)}$ \\
0.1 & -- & 0.5   & $0.46 \pm  0.11\,\mathrm{(stat.)}\,^{+0.02}_{-0.06}\,\mathrm{(syst.)}$ \\
0.5 & -- & 0.9   & $0.45 \pm  0.09\,\mathrm{(stat.)}\,^{+0.04}_{-0.07}\,\mathrm{(syst.)}$ \\

\hline
\multicolumn{3}{|l|}{$\zpom<0.9$}   & \\
\hline
--0.7 & -- & --\,0.3   & $0.70 \pm  0.11\,\mathrm{(stat.)}\,^{+0.04}_{-0.08}\,\mathrm{(syst.)}$ \\
--\,0.3 & -- & 0.1   & $0.47 \pm  0.09\,\mathrm{(stat.)}\,^{+0.04}_{-0.06}\,\mathrm{(syst.)}$ \\
0.1 & -- & 0.5   & $0.28 \pm  0.07\,\mathrm{(stat.)}\,^{+0.02}_{-0.03}\,\mathrm{(syst.)}$ \\
0.5 & -- & 0.9   & $0.26 \pm  0.07\,\mathrm{(stat.)}\,^{+0.02}_{-0.04}\,\mathrm{(syst.)}$ \\
\hline
\multicolumn{3}{|l|}{$\zpom\geq 0.9$}   & \\
\hline

--0.7 & -- & --\,0.3   & $0.44 \pm  0.11\,\mathrm{(stat.)}\,^{+0.11}_{-0.09}\,\mathrm{(syst.)}$ \\
--\,0.3 & -- & 0.1   & $0.29 \pm  0.09\,\mathrm{(stat.)}\,^{+0.07}_{-0.06}\,\mathrm{(syst.)}$ \\
0.1 & -- & 0.5   & $0.21 \pm  0.07\,\mathrm{(stat.)}\,^{+0.04}_{-0.05}\,\mathrm{(syst.)}$ \\
0.5 & -- & 0.9   & $0.19 \pm  0.07\,\mathrm{(stat.)}\,^{+0.03}_{-0.05}\,\mathrm{(syst.)}$ \\
\hline
\end{tabular}
\end{center}
\caption{Differential cross-section
$\frac{d\sigma}{d\eta^{\gamma}}$ for photons accompanied by at least one jet in 
diffractive photoproduction. (Figure~\ref{fig:xsgam}(d--f))
}
\label{tab:etagj}
\end{table}
\begin{table}
\begin{center}
\begin{tabular}{|rcr|c|}
\hline
\multicolumn{3}{|c|}{$W$ range}   &       \\[-1.8ex]
\multicolumn{3}{|c|}{ (GeV)}   &  \raisebox{1.8ex}{$\frac{d\sigma}{dW}$ ($\mathrm{pb}\,\mathrm{GeV}^{-1})$}   \\ 
\hline\hline
\multicolumn{3}{|l|}{$0<\zpom\leq 1.0$}   & \\
\hline

140 & -- & 160   & $0.0089 \pm  0.0020\,\mathrm{(stat.)}\,^{+0.0005}_{-0.0012}\,\mathrm{(syst.)}$ \\
160 & -- & 180   & $0.0163 \pm  0.0027\,\mathrm{(stat.)}\,^{+0.0013}_{-0.0023}\,\mathrm{(syst.)}$ \\
180 & -- & 200   & $0.0121 \pm  0.0023\,\mathrm{(stat.)}\,^{+0.0007}_{-0.0015}\,\mathrm{(syst.)}$ \\
200 & -- & 220   & $0.0102 \pm  0.0024\,\mathrm{(stat.)}\,^{+0.0008}_{-0.0012}\,\mathrm{(syst.)}$ \\
220 & -- & 240   & $0.0059 \pm  0.0015\,\mathrm{(stat.)}\,^{+0.0005}_{-0.0007}\,\mathrm{(syst.)}$ \\
240 & -- & 260   & $0.0050 \pm  0.0015\,\mathrm{(stat.)}\,^{+0.0002}_{-0.0006}\,\mathrm{(syst.)}$ \\
\hline
\multicolumn{3}{|l|}{$\zpom<0.9$}   & \\
\hline

140 & -- & 160   & $0.0045 \pm  0.0012\,\mathrm{(stat.)}\,^{+0.0002}_{-0.0005}\,\mathrm{(syst.)}$ \\
160 & -- & 180   & $0.0086 \pm  0.0016\,\mathrm{(stat.)}\,^{+0.0003}_{-0.0012}\,\mathrm{(syst.)}$ \\
180 & -- & 200   & $0.0079 \pm  0.0016\,\mathrm{(stat.)}\,^{+0.0004}_{-0.0009}\,\mathrm{(syst.)}$ \\
200 & -- & 220   & $0.0054 \pm  0.0015\,\mathrm{(stat.)}\,^{+0.0003}_{-0.0006}\,\mathrm{(syst.)}$ \\
220 & -- & 240   & $0.0044 \pm  0.0013\,\mathrm{(stat.)}\,^{+0.0003}_{-0.0005}\,\mathrm{(syst.)}$ \\
240 & -- & 260   & $0.0032 \pm  0.0013\,\mathrm{(stat.)}\,^{+0.0003}_{-0.0004}\,\mathrm{(syst.)}$ \\

\hline
\multicolumn{3}{|l|}{$\zpom \geq0.9$}   & \\
\hline

140 & -- & 160   & $0.0042 \pm  0.0016\,\mathrm{(stat.)}\,^{+0.0002}_{-0.0008}\,\mathrm{(syst.)}$ \\
160 & -- & 180   & $0.0062 \pm  0.0016\,\mathrm{(stat.)}\,^{+0.0010}_{-0.0010}\,\mathrm{(syst.)}$ \\
180 & -- & 200   & $0.0031 \pm  0.0013\,\mathrm{(stat.)}\,^{+0.0012}_{-0.0007}\,\mathrm{(syst.)}$ \\
200 & -- & 220   & $0.0047 \pm  0.0017\,\mathrm{(stat.)}\,^{+0.0015}_{-0.0012}\,\mathrm{(syst.)}$ \\
220 & -- & 240   & $0.0015 \pm  0.0008\,\mathrm{(stat.)}\,^{+0.0005}_{-0.0004}\,\mathrm{(syst.)}$ \\
240 & -- & 260   & $0.0015 \pm  0.0007\,\mathrm{(stat.)}\,^{+0.0003}_{-0.0004}\,\mathrm{(syst.)}$ \\

\hline
\end{tabular}
\end{center}
\caption{Differential cross-section
$\frac{d\sigma}{dW}$ for photons accompanied by at least one jet
in diffractive photoproduction. 
 (Figure~\ref{fig:xsgam}(g--i))
}
\label{tab:Wj}
\end{table}

\begin{table}
\begin{center}
\begin{tabular}{|rcr|c|}
\hline
\multicolumn{3}{|c|}{$E^{\jet}_T  $ range}   &       \\[-1.8ex]
\multicolumn{3}{|c|}{ (GeV)}   &  \raisebox{1.8ex}{$\frac{d\sigma}{dE^{\jet}_T  }$ ($\mathrm{pb}\,\mathrm{GeV}^{-1})$}   \\ 
\hline\hline
\multicolumn{3}{|l|}{$0<\zpom\leq 1.0$}   & \\
\hline

4.0 & -- & 6.0   & $0.178\pm 0.032\,\mathrm{(stat.)}\,^{+0.011}_{-0.029}\,\mathrm{(syst.)}$   \\
6.0 & -- & 8.0   & $0.253\pm 0.036\,\mathrm{(stat.)}\,^{+0.023}_{-0.030}\,\mathrm{(syst.)}$   \\
8.0 & -- & 10.0   & $0.112\pm0.019 \,\mathrm{(stat.)}\,^{+0.007}_{-0.015}\,\mathrm{(syst.)}$  \\
10.0 & -- & 15.0   & $0.016\pm0.004 \,\mathrm{(stat.)}\,^{+0.003}_{-0.002}\,\mathrm{(syst.)}$ \\
\hline
\multicolumn{3}{|l|}{$\zpom <0.9$}   & \\
\hline
4.0 & -- & 6.0   & $0.136\pm  0.025\,\mathrm{(stat.)}\,^{+0.007}_{-0.020}\,\mathrm{(syst.)}$  \\
6.0 & -- & 8.0   & $0.128\pm  0.022\,\mathrm{(stat.)}\,^{+0.016}_{-0.015}\,\mathrm{(syst.)}$  \\
8.0 & -- & 10.0   & $0.061\pm  0.012\,\mathrm{(stat.)}\,^{+0.003}_{-0.006}\,\mathrm{(syst.)}$ \\
10.0 & -- & 15.0   & $0.006\pm  0.002\,\mathrm{(stat.)}\,^{+0.001}_{-0.001}\,\mathrm{(syst.)}$\\

\hline
\multicolumn{3}{|l|}{$\zpom \geq0.9$}   & \\
\hline

4.0 & -- & 6.0   & $0.030\pm  0.013\,\mathrm{(stat.)}\,^{+0.010}_{-0.007}\,\mathrm{(syst.)}$  \\
6.0 & -- & 8.0   & $0.126\pm  0.026\,\mathrm{(stat.)}\,^{+0.024}_{-0.024}\,\mathrm{(syst.)}$  \\
8.0 & -- & 10.0   & $0.043\pm  0.013\,\mathrm{(stat.)}\,^{+0.006}_{-0.010}\,\mathrm{(syst.)}$ \\
10.0 & -- & 15.0   & $0.010\pm  0.003\,\mathrm{(stat.)}\,^{+0.002}_{-0.002}\,\mathrm{(syst.)}$\\

\hline
\end{tabular}
\end{center}

\caption{Differential cross-section
$\frac{d\sigma}{dE^{\jet}_T}$ for photons accompanied by at least one jet
in diffractive photoproduction. (Figure~\ref{fig:xsjet}(a--c))
}                                         
\label{tab:etjj}
\end{table}
\begin{table}
\begin{center}
\begin{tabular}{|rcr|c|}
\hline
\multicolumn{3}{|c|}{ $\eta^{\jet  }$ range } &\multicolumn{1}{|c|}{ $\frac{d\sigma}{d\eta^{\jet  }}$ 
($\mathrm{pb}$)}     \\[0.5mm]
\hline\hline
\multicolumn{3}{|l|}{$0<\zpom\leq 1.0$}   & \\

\hline
--1.5 & -- & --\,0.7   & $0.38 \pm  0.06\,\mathrm{(stat.)}\,^{+0.03}_{-0.05}\,\mathrm{(syst.)}$ \\
--\,0.7 & -- & 0.1   & $0.53 \pm  0.08\,\mathrm{(stat.)}\,^{+0.04}_{-0.06}\,\mathrm{(syst.)}$ \\
0.1 & -- & 0.9   & $0.43 \pm  0.07\,\mathrm{(stat.)}\,^{+0.02}_{-0.07}\,\mathrm{(syst.)}$ \\
0.9 & -- & 1.8   & $0.09 \pm  0.03\,\mathrm{(stat.)}\,^{+0.00}_{-0.01}\,\mathrm{(syst.)}$ \\

\hline
\multicolumn{3}{|l|}{$\zpom<0.9$}   & \\
\hline

--1.5 & -- & --\,0.7   & $0.27 \pm  0.05\,\mathrm{(stat.)}\,^{+0.02}_{-0.03}\,\mathrm{(syst.)}$ \\
--\,0.7 & -- & 0.1   & $0.32 \pm  0.05\,\mathrm{(stat.)}\,^{+0.02}_{-0.03}\,\mathrm{(syst.)}$ \\
0.1 & -- & 0.9   & $0.24 \pm  0.05\,\mathrm{(stat.)}\,^{+0.01}_{-0.04}\,\mathrm{(syst.)}$ \\
0.9 & -- & 1.8   & $0.03 \pm  0.02\,\mathrm{(stat.)}\,^{+0.01}_{-0.01}\,\mathrm{(syst.)}$ \\
\hline
\multicolumn{3}{|l|}{$\zpom\geq 0.9$}   & \\
\hline

--1.5 & -- & --\,0.7   & $0.08 \pm  0.03\,\mathrm{(stat.)}\,^{+0.03}_{-0.02}\,\mathrm{(syst.)}$ \\
--\,0.7 & -- & 0.1   & $0.18 \pm  0.05\,\mathrm{(stat.)}\,^{+0.04}_{-0.04}\,\mathrm{(syst.)}$ \\
0.1 & -- & 0.9   & $0.17 \pm  0.04\,\mathrm{(stat.)}\,^{+0.03}_{-0.04}\,\mathrm{(syst.)}$ \\
0.9 & -- & 1.8   & $0.09 \pm  0.03\,\mathrm{(stat.)}\,^{+0.03}_{-0.02}\,\mathrm{(syst.)}$ \\

\hline
\end{tabular}
\end{center}
\caption{Differential cross-section
$\frac{d\sigma}{d\eta^{\jet  }}$ for photons accompanied by at least one jet in diffractive photoproduction. (Figure~\ref{fig:xsjet}(d--f))
}
\label{tab:etajj}
\end{table}

\begin{table}
\begin{center}
\begin{tabular}{|rcr|c|}
\hline
\multicolumn{3}{|c|}{ $E^{\gamma}_T / E^{\jet}_T $  range } 
&\multicolumn{1}{|c|}{ $\frac{d\sigma}{d(E^{\gamma}_T / E^{\jet}_T)}$
($\mathrm{pb}$)}     \\[0.5mm]
\hline\hline
\multicolumn{3}{|l|}{$0<\zpom\leq 1.0$}   & \\

\hline
0.4 & -- & 0.6   & $0.012\pm  0.007\,\mathrm{(stat.)}\,^{+0.005}_{-0.002}\,\mathrm{(syst.)}$ \\
0.6 & -- & 0.8   & $0.62 \pm  0.13\,\mathrm{(stat.)}\,^{+0.08 }_{-0.15 }\,\mathrm{(syst.)}$ \\
0.8 & -- & 1.0   & $2.68 \pm  0.45\,\mathrm{(stat.)}\,^{+0.43 }_{-0.30 }\,\mathrm{(syst.)}$ \\
1.0 & -- & 1.2   & $1.29 \pm  0.24 \,\mathrm{(stat.)}\,^{+0.13 }_{-0.20 }\,\mathrm{(syst.)}$ \\
1.2 & -- & 1.4   & $0.35 \pm  0.09\,\mathrm{(stat.)}\,^{+0.11 }_{-0.09 }\,\mathrm{(syst.)}$ \\
1.4 & -- & 1.6   & $0.15 \pm  0.05\,\mathrm{(stat.)}\,^{+0.05 }_{-0.03 }\,\mathrm{(syst.)}$ \\

\hline
\multicolumn{3}{|l|}{$\zpom<0.9$}   & \\
\hline
 
0.4 & -- & 0.6   & $0.010\pm 0.006\,\mathrm{(stat.)}\,^{+0.005}_{-0.002}\,\mathrm{(syst.)}$ \\
0.6 & -- & 0.8   & $0.36 \pm  0.10\,\mathrm{(stat.)}\,^{+0.05 }_{-0.08 }\,\mathrm{(syst.)}$ \\
0.8 & -- & 1.0   & $1.28 \pm  0.27\,\mathrm{(stat.)}\,^{+0.24 }_{-0.20 }\,\mathrm{(syst.)}$ \\
1.0 & -- & 1.2   & $0.95 \pm  0.22\,\mathrm{(stat.)}\,^{+0.07 }_{-0.17 }\,\mathrm{(syst.)}$ \\
1.2 & -- & 1.4   & $0.31 \pm  0.09\,\mathrm{(stat.)}\,^{+0.08 }_{-0.07 }\,\mathrm{(syst.)}$ \\
1.4 & -- & 1.6   & $0.15 \pm  0.05\,\mathrm{(stat.)}\,^{+0.05 }_{-0.02 }\,\mathrm{(syst.)}$ \\
\hline
\multicolumn{3}{|l|}{$\zpom\geq 0.9$}   & \\
\hline

0.6 & -- & 0.8   & $0.24 \pm  0.08\,\mathrm{(stat.)}\,^{+0.07}_{-0.07}\,\mathrm{(syst.)}$ \\
0.8 & -- & 1.0   & $1.19 \pm  0.29\,\mathrm{(stat.)}\,^{+0.27}_{-0.16}\,\mathrm{(syst.)}$ \\
1.0 & -- & 1.2   & $0.33 \pm  0.09\,\mathrm{(stat.)}\,^{+0.16}_{-0.06}\,\mathrm{(syst.)}$ \\
\hline
\end{tabular}
\end{center}
\caption{Differential cross-section
$\frac{d\sigma}{d( E^{\gamma}_T / E^{\jet}_T) }$
for photons accompanied by at least one jet in diffractive photoproduction.
Omitted values for $\zpom \geq0.9$
 are consistent with zero. (Figure~\ref{fig:xsjet}(g--i))
}
\label{tab:etrat}
\end{table}
\begin{table}
\begin{center}
\begin{tabular}{|rcr|c|}
\hline
\multicolumn{3}{|c|}{ $\xgamm $ range } &\multicolumn{1}{|c|}{ $\frac{d\sigma}{d\xgamm}$
($\mathrm{pb}$)}     \\[0.5mm]
\hline\hline
\multicolumn{3}{|l|}{$0<\zpom\leq 1.0$}   & \\

\hline
0.1 & -- & 0.6   & $0.16 \pm  0.08\,\mathrm{(stat.)}\,^{+0.03}_{-0.05}\,\mathrm{(syst.)}$ \\
0.6 & -- & 0.7   & $0.54 \pm  0.20\,\mathrm{(stat.)}\,^{+0.09}_{-0.11}\,\mathrm{(syst.)}$ \\
0.7 & -- & 0.8   & $1.25 \pm  0.31\,\mathrm{(stat.)}\,^{+0.09}_{-0.20}\,\mathrm{(syst.)}$ \\
0.8 & -- & 0.9   & $1.95 \pm  0.35\,\mathrm{(stat.)}\,^{+0.18}_{-0.20}\,\mathrm{(syst.)}$ \\
0.9 & -- & 1.0   & $5.98 \pm  0.64\,\mathrm{(stat.)}\,^{+0.50}_{-0.81}\,\mathrm{(syst.)}$ \\

\hline
\multicolumn{3}{|l|}{$\zpom<0.9$}   & \\
\hline

0.1 & -- & 0.6   & $0.08 \pm  0.07\,\mathrm{(stat.)}\,^{+0.04}_{-0.04}\,\mathrm{(syst.)}$ \\
0.6 & -- & 0.7   & $0.49 \pm  0.18\,\mathrm{(stat.)}\,^{+0.05}_{-0.10}\,\mathrm{(syst.)}$ \\
0.7 & -- & 0.8   & $1.01 \pm  0.27\,\mathrm{(stat.)}\,^{+0.07}_{-0.16}\,\mathrm{(syst.)}$ \\
0.8 & -- & 0.9   & $1.80 \pm  0.34\,\mathrm{(stat.)}\,^{+0.27}_{-0.21}\,\mathrm{(syst.)}$ \\
0.9 & -- & 1.0   & $2.81 \pm  0.37\,\mathrm{(stat.)}\,^{+0.09}_{-0.30}\,\mathrm{(syst.)}$ \\

\hline
\multicolumn{3}{|l|}{$\zpom\geq 0.9$}   & \\
\hline

0.1 & -- & 0.6   & $0.11 \pm  0.05\,\mathrm{(stat.)}\,^{+0.02}_{-0.05}\,\mathrm{(syst.)}$ \\
0.6 & -- & 0.7   & $0.08 \pm  0.07\,\mathrm{(stat.)}\,^{+0.05}_{-0.02}\,\mathrm{(syst.)}$ \\
0.7 & -- & 0.8   & $0.21 \pm  0.16\,\mathrm{(stat.)}\,^{+0.03}_{-0.04}\,\mathrm{(syst.)}$ \\
0.8 & -- & 0.9   & $0.26 \pm  0.13\,\mathrm{(stat.)}\,^{+0.06}_{-0.07}\,\mathrm{(syst.)}$ \\
0.9 & -- & 1.0   & $2.78 \pm  0.48\,\mathrm{(stat.)}\,^{+0.57}_{-0.54}\,\mathrm{(syst.)}$ \\

\hline
\end{tabular}
\end{center}
\caption{Differential cross-section
$\frac{d\sigma}{d\xgamm }$ for photons accompanied by at least one jet in diffractive photoproduction.
 (Figure~\ref{fig:xsxm}(a--c))
}\label{tab:xg}
\end{table}
\begin{table}
\begin{center}
\begin{tabular}{|rcr|c|}
\hline
\multicolumn{3}{|c|}{ $\xpom$ range }  &  $\frac{d\sigma}{d \xpom}$ ($\mathrm{pb}$) \\[0.5mm]
\hline\hline
\multicolumn{3}{|l|}{$0<\zpom\leq 1.0$}   & \\

\hline
0.0 &--& 0.005   & $26.0 \pm  7.0\,\mathrm{(stat.)}\,^{+3.3}_{-3.1}\,\mathrm{(syst.)}$ \\
0.005 &--& 0.01   & $76.4 \pm  11.8\,\mathrm{(stat.)}\,^{+5.6}_{-9.1 }\,\mathrm{(syst.)}$ \\
0.01 &--& 0.015   & $70.6 \pm  10.2\,\mathrm{(stat.)}\,^{+3.2}_{-11.0}\,\mathrm{(syst.)}$ \\
0.015 &--&   0.02 & $37.8 \pm  7.9\,\mathrm{(stat.)}\,^{+2.6}_{-4.8}\,\mathrm{(syst.)}$ \\
0.02 &--&   0.025   & $11.7 \pm  4.1\,\mathrm{(stat.)}\,^{+0.8}_{-1.4}\,\mathrm{(syst.)}$ \\
0.025 &--&   0.03   & $5.4 \pm  3.3\,\mathrm{(stat.)}\,^{+2.0}_{-0.6}\,\mathrm{(syst.)}$ \\

\hline

\multicolumn{3}{|l|}{$\zpom<0.9$}   & \\
\hline
0.0 &--&     0.005   & $ 9.0 \pm  3.0\,\mathrm{(stat.)}\,^{+4.3}_{-1.5}\,\mathrm{(syst.)}$ \\
0.005 &--& 0.01   & $40.4 \pm   7.5\,\mathrm{(stat.)}\,^{+3.7}_{-4.5 }\,\mathrm{(syst.)}$ \\
0.01 &--& 0.015   & $49.8 \pm   8.5\,\mathrm{(stat.)}\,^{+1.4}_{-7.8 }\,\mathrm{(syst.)}$ \\
0.015 &--&   0.02   & $26.6 \pm  6.5\,\mathrm{(stat.)}\,^{+1.4}_{-3.4}\,\mathrm{(syst.)}$ \\
0.02 &--&   0.025   & $9.5  \pm  3.6\,\mathrm{(stat.)}\,^{+0.7}_{-1.2}\,\mathrm{(syst.)}$ \\
0.025 &--&  .0.03   & $2.3 \pm  2.0\,\mathrm{(stat.)}\,^{+1.1}_{-0.2}\,\mathrm{(syst.)}$ \\

\hline
\multicolumn{3}{|l|}{$\zpom\geq 0.9$}   & \\
\hline
0.0 &--&     0.005   & $11.6 \pm  5.5\,\mathrm{(stat.)}\,^{+4.7}_{-3.6}\,\mathrm{(syst.)}$ \\
0.005 &--& 0.01   & $31.9 \pm   8.9\,\mathrm{(stat.)}\,^{+9.2}_{-7.1 }\,\mathrm{(syst.)}$ \\
0.01 &--& 0.015   & $20.8 \pm   5.0\,\mathrm{(stat.)}\,^{+4.0}_{-4.8 }\,\mathrm{(syst.)}$ \\
0.015 &--&   0.02   & $10.6 \pm  3.7\,\mathrm{(stat.)}\,^{+1.9}_{-1.7}\,\mathrm{(syst.)}$ \\
0.02 &--&   0.025   & $ 3.1 \pm  3.3\,\mathrm{(stat.)}\,^{+0.4}_{-0.8}\,\mathrm{(syst.)}$ \\
0.025 &--&   0.03   & $9.5 \pm 10.5\,\mathrm{(stat.)}\,^{+2.3}_{-4.7}\,\mathrm{(syst.)}$ \\

\hline
\end{tabular}
\end{center}
\caption{Differential cross-section
$\frac{d\sigma}{d\xpom  }$ for photons accompanied by at least one jet in diffractive photoproduction. 
 (Figure~\ref{fig:xsxm}(d--f))}
\label{tab:xpj}
\end{table}


\begin{table}
\begin{center}
\begin{tabular}{|rcr|c|}
\hline
\multicolumn{3}{|c|}{$M_X$ range}   &       \\[-1.8ex]
\multicolumn{3}{|c|}{ (GeV)}   &  \raisebox{1.8ex}{$\frac{d\sigma}{dM_X}$ ($\mathrm{pb}\,\mathrm{GeV}^{-1})$}   \\ 
\hline\hline
\multicolumn{3}{|l|}{$0<\zpom\leq 1.0$}   & \\
\hline

10.0 &--& 15.0   & $0.042\pm 0.008\,\mathrm{(stat.)}\,^{+0.003}_{-0.005}\,\mathrm{(syst.)}$ \\
15.0 &--& 20.0   & $0.091\pm 0.012\,\mathrm{(stat.)}\,^{+0.007}_{-0.013}\,\mathrm{(syst.)}$ \\
20.0 &--& 25.0   & $0.055\pm  0.009\,\mathrm{(stat.)}\,^{+0.007}_{-0.009}\,\mathrm{(syst.)}$ \\
25.0 &--& 30.0   & $0.029\pm0.006\,\mathrm{(stat.)}\,^{+0.001}_{-0.004}\,\mathrm{(syst.)}$ \\
30.0 &--& 40.0   & $0.004\pm0.002\,\mathrm{(stat.)}\,^{+0.001}_{-0.000}\,\mathrm{(syst.)}$ \\

\hline

\multicolumn{3}{|l|}{$\zpom <0.9$}   & \\
\hline
10.0 &--& 15.0   & $0.013\pm 0.003\,\mathrm{(stat.)}\,^{+0.003}_{-0.002}\,\mathrm{(syst.)}$ \\
15.0 &--& 20.0   & $0.051\pm 0.009\,\mathrm{(stat.)}\,^{+0.003}_{-0.007}\,\mathrm{(syst.)}$ \\
20.0 &--& 25.0   & $0.042\pm  0.008\,\mathrm{(stat.)}\,^{+0.007}_{-0.007}\,\mathrm{(syst.)}$ \\
25.0 &--& 30.0   & $0.024\pm0.006\,\mathrm{(stat.)}\,^{+0.001}_{-0.004}\,\mathrm{(syst.)}$ \\
30.0 &--& 40.0   & $0.003\pm0.002\,\mathrm{(stat.)}\,^{+0.001}_{-0.000}\,\mathrm{(syst.)}$ \\

\hline

\multicolumn{3}{|l|}{$\zpom \geq0.9$}   & \\
\hline

10.0 &--& 15.0   & $0.024\pm 0.007\,\mathrm{(stat.)}\,^{+0.004}_{-0.005}\,\mathrm{(syst.)}$ \\
15.0 &--& 20.0   & $0.039\pm 0.008\,\mathrm{(stat.)}\,^{+0.008}_{-0.009}\,\mathrm{(syst.)}$ \\
20.0 &--& 25.0   & $0.014\pm  0.004\,\mathrm{(stat.)}\,^{+0.004}_{-0.003}\,\mathrm{(syst.)}$ \\
25.0 &--& 30.0   & $0.005\pm0.004\,\mathrm{(stat.)}\,^{+0.001}_{-0.001}\,\mathrm{(syst.)}$ \\
30.0 &--& 40.0   & $0.002\pm0.002\,\mathrm{(stat.)}\,^{+0.001}_{-0.000}\,\mathrm{(syst.)}$ \\

\hline
\end{tabular}
\end{center}
\caption{Differential cross-section
$\frac{d\sigma}{dM_X}$ for photons accompanied by at least one jet
in diffractive photoproduction.
 (Figure~\ref{fig:xsxm}(g--i))
}
\label{tab:mxj}
\end{table}


\begin{table}
\begin{center}
\begin{tabular}{|rcr|c|}
\hline
\multicolumn{3}{|c|}{$\Delta\phi$ range}   &       \\[-1.8ex]
\multicolumn{3}{|c|}{ (deg.)}   &  \raisebox{1.8ex}{$\frac{d\sigma}{d\Delta\phi}$ ($\mathrm{pb}\,\mathrm{deg.}^{-1})$} \\ 
\hline\hline
\multicolumn{3}{|l|}{$0<\zpom\leq 1.0$}   & \\
\hline
130 & -- & 140   & $0.001 \pm  0.001\,\mathrm{(stat.)}\,^{+0.000}_{-0.000}\,\mathrm{(syst.)}$ \\
140 & -- & 150   & $0.002 \pm  0.002\,\mathrm{(stat.)}\,^{+0.001}_{-0.001}\,\mathrm{(syst.)}$ \\
150 & -- & 160   & $0.007 \pm  0.002\,\mathrm{(stat.)}\,^{+0.002}_{-0.001}\,\mathrm{(syst.)}$ \\
160 & -- & 170   & $0.017 \pm  0.003\,\mathrm{(stat.)}\,^{+0.002}_{-0.003}\,\mathrm{(syst.)}$ \\
170 & -- & 180   & $0.077 \pm  0.009\,\mathrm{(stat.)}\,^{+0.004}_{-0.009}\,\mathrm{(syst.)}$ \\
\hline
\multicolumn{3}{|l|}{$\zpom<0.9$}   & \\
\hline
130 & -- & 140   & $0.001 \pm  0.001\,\mathrm{(stat.)}\,^{+0.000}_{-0.000}\,\mathrm{(syst.)}$ \\
140 & -- & 150   & $0.002 \pm  0.002\,\mathrm{(stat.)}\,^{+0.001}_{-0.001}\,\mathrm{(syst.)}$ \\
150 & -- & 160   & $0.006 \pm  0.002\,\mathrm{(stat.)}\,^{+0.002}_{-0.001}\,\mathrm{(syst.)}$ \\
160 & -- & 170   & $0.015 \pm  0.003\,\mathrm{(stat.)}\,^{+0.001}_{-0.002}\,\mathrm{(syst.)}$ \\
170 & -- & 180   & $0.038 \pm  0.005\,\mathrm{(stat.)}\,^{+0.002}_{-0.004}\,\mathrm{(syst.)}$ \\
\hline
\multicolumn{3}{|l|}{$\zpom \geq0.9$}   & \\
\hline
160 & -- & 170   & $0.003 \pm  0.001\,\mathrm{(stat.)}\,^{+0.002}_{-0.001}\,\mathrm{(syst.)}$ \\
170 & -- & 180   & $0.037 \pm  0.006\,\mathrm{(stat.)}\,^{+0.007}_{-0.006}\,\mathrm{(syst.)}$ \\

\hline
\end{tabular}
\end{center}
\caption{Differential cross-section
$\frac{d\sigma}{d\Delta\phi}$ for photons accompanied by at least one jet
in diffractive photoproduction, where 
$\delphi = |\phi^\gamma -\phi^{\mathrm jet}|$.
 Omitted values for $\zpom \geq0.9$ are consistent with zero.
 (Figure~\ref{fig:xsphe}(a--c))
}
\label{tab:phi}
\end{table}

\begin{table}
\begin{center}
\begin{tabular}{|rcr|c|}
\hline
\multicolumn{3}{|c|}{ $\Delta\eta$ range }  &  $\frac{d\sigma}{d \Delta\eta}$ ($\mathrm{pb}$) \\[0.5mm]
\hline\hline
\multicolumn{3}{|l|}{$0<\zpom\leq 1.0$}   & \\
\hline

--2.9 & -- & --2.2   & $0.006 \pm  0.005\,\mathrm{(stat.)}\,^{+0.002}_{-0.001}\,\mathrm{(syst.)}$ \\
--2.2 & -- & --1.5   & $0.018 \pm  0.028\,\mathrm{(stat.)}\,^{+0.007}_{-0.007}\,\mathrm{(syst.)}$ \\
--1.5 & -- & --0.8   & $0.187 \pm  0.053\,\mathrm{(stat.)}\,^{+0.007}_{-0.042}\,\mathrm{(syst.)}$ \\
--0.8 & -- & --0.1   & $0.438 \pm  0.074\,\mathrm{(stat.)}\,^{+0.029}_{-0.057}\,\mathrm{(syst.)}$ \\
--0.1 & -- & 0.6     & $0.541 \pm  0.088\,\mathrm{(stat.)}\,^{+0.047}_{-0.067}\,\mathrm{(syst.)}$ \\ 
  0.6 & -- & 1.3     & $0.329 \pm  0.062\,\mathrm{(stat.)}\,^{+0.012}_{-0.044}\,\mathrm{(syst.)}$ \\
  1.3 & -- & 2.0     & $0.115 \pm  0.032\,\mathrm{(stat.)}\,^{+0.024}_{-0.013}\,\mathrm{(syst.)}$ \\
  2.0 & -- & 2.7     & $0.015 \pm  0.014\,\mathrm{(stat.)}\,^{+0.001}_{-0.002}\,\mathrm{(syst.)}$ \\
\hline
\multicolumn{3}{|l|}{$\zpom<0.9$}   & \\
\hline

--2.9 & -- & --1.5   & $0.000 \pm  0.004\,\mathrm{(stat.)}\,^{+0.000}_{-0.000}\,\mathrm{(syst.)}$ \\
--1.5 & -- & --0.8   & $0.085 \pm  0.035\,\mathrm{(stat.)}\,^{+0.002}_{-0.020}\,\mathrm{(syst.)}$ \\
--0.8 & -- & --0.1   & $0.242 \pm  0.047\,\mathrm{(stat.)}\,^{+0.018}_{-0.029}\,\mathrm{(syst.)}$ \\
--0.1 & -- & 0.6     & $0.343 \pm  0.058\,\mathrm{(stat.)}\,^{+0.021}_{-0.039}\,\mathrm{(syst.)}$ \\ 
  0.6 & -- & 1.3     & $0.220 \pm  0.048\,\mathrm{(stat.)}\,^{+0.017}_{-0.027}\,\mathrm{(syst.)}$ \\
  1.3 & -- & 2.7     & $0.040 \pm  0.012\,\mathrm{(stat.)}\,^{+0.008}_{-0.005}\,\mathrm{(syst.)}$ \\

\hline
\multicolumn{3}{|l|}{$\zpom \geq0.9$}   & \\
\hline

--2.9 & -- & --1.5   & $0.028 \pm  0.013\,\mathrm{(stat.)}\,^{+0.010}_{-0.003}\,\mathrm{(syst.)}$ \\
--1.5 & -- & --0.8   & $0.106 \pm  0.029\,\mathrm{(stat.)}\,^{+0.025}_{-0.028}\,\mathrm{(syst.)}$ \\
--0.8 & -- & --0.1   & $0.188 \pm  0.057\,\mathrm{(stat.)}\,^{+0.035}_{-0.042}\,\mathrm{(syst.)}$ \\
--0.1 & -- & 0.6     & $0.144 \pm  0.045\,\mathrm{(stat.)}\,^{+0.029}_{-0.030}\,\mathrm{(syst.)}$ \\ 
  0.6 & -- & 1.3     & $0.097 \pm  0.037\,\mathrm{(stat.)}\,^{+0.026}_{-0.027}\,\mathrm{(syst.)}$ \\
  1.3 & -- & 2.7     & $0.015 \pm  0.015\,\mathrm{(stat.)}\,^{+0.007}_{-0.011}\,\mathrm{(syst.)}$ \\

\hline
\end{tabular}
\end{center}
\caption{Differential cross-section
$\frac{d\sigma}{d\Delta\eta}$ for photons accompanied by at least one jet
in diffractive photoproduction, where 
$\Deleta = \eta^\gamma - \etajet$.
 (Figure~\ref{fig:xsphe}(d--f))
}
\label{tab:deta}
\end{table}


\begin{table}
\begin{center}
\begin{tabular}{|rcr|c|}
\hline
\multicolumn{3}{|c|}{ $\etamax   $ range }  &  $\frac{d\sigma}{d \etamax}$ ($\mathrm{pb}$) \\[0.5mm]
\hline\hline
\multicolumn{3}{|l|}{$0<\zpom\leq 1.0$}   & \\
\hline

--1.0 & -- & 0.0     & $0.091 \pm  0.023\,\mathrm{(stat.)}\,^{+0.014}_{-0.101}\,\mathrm{(syst.)}$ \\
  0.0 & -- & 0.5     & $0.279 \pm  0.064\,\mathrm{(stat.)}\,^{+0.010}_{-0.042}\,\mathrm{(syst.)}$ \\
  0.5 & -- & 1.0     & $0.282 \pm  0.084\,\mathrm{(stat.)}\,^{+0.039}_{-0.034}\,\mathrm{(syst.)}$ \\
  1.0 & -- & 1.5     & $0.537 \pm  0.091\,\mathrm{(stat.)}\,^{+0.034}_{-0.069}\,\mathrm{(syst.)}$ \\
  1.5 & -- & 2.0     & $0.433 \pm  0.089\,\mathrm{(stat.)}\,^{+0.020}_{-0.062}\,\mathrm{(syst.)}$ \\
  2.0 & -- & 2.5     & $0.543 \pm  0.105\,\mathrm{(stat.)}\,^{+0.051}_{-0.069}\,\mathrm{(syst.)}$ \\
\hline
\multicolumn{3}{|l|}{$\zpom<0.9$}   & \\
\hline

--1.0 & -- & 0.0     & $0.014 \pm  0.005\,\mathrm{(stat.)}\,^{+0.009}_{-0.004}\,\mathrm{(syst.)}$ \\
  0.0 & -- & 0.5     & $0.049 \pm  0.018\,\mathrm{(stat.)}\,^{+0.012}_{-0.011}\,\mathrm{(syst.)}$ \\
  0.5 & -- & 1.0     & $0.104 \pm  0.040\,\mathrm{(stat.)}\,^{+0.017}_{-0.011}\,\mathrm{(syst.)}$ \\
  1.0 & -- & 1.5     & $0.318 \pm  0.065\,\mathrm{(stat.)}\,^{+0.008}_{-0.036}\,\mathrm{(syst.)}$ \\
  1.5 & -- & 2.0     & $0.375 \pm  0.078\,\mathrm{(stat.)}\,^{+0.009}_{-0.051}\,\mathrm{(syst.)}$ \\
  2.0 & -- & 2.5     & $0.464 \pm  0.095\,\mathrm{(stat.)}\,^{+0.046}_{-0.056}\,\mathrm{(syst.)}$ \\

\hline
\multicolumn{3}{|l|}{$\zpom \geq0.9$}   & \\
\hline

--1.0 & -- & 0.0     & $0.054 \pm  0.023\,\mathrm{(stat.)}\,^{+0.024}_{-0.014}\,\mathrm{(syst.)}$ \\
  0.0 & -- & 0.5     & $0.201 \pm  0.059\,\mathrm{(stat.)}\,^{+0.040}_{-0.042}\,\mathrm{(syst.)}$ \\
  0.5 & -- & 1.0     & $0.136 \pm  0.064\,\mathrm{(stat.)}\,^{+0.040}_{-0.039}\,\mathrm{(syst.)}$ \\
  1.0 & -- & 1.5     & $0.159 \pm  0.053\,\mathrm{(stat.)}\,^{+0.035}_{-0.034}\,\mathrm{(syst.)}$ \\
  1.5 & -- & 2.0     & $0.048  \pm  0.040 \,\mathrm{(stat.)}\,^{+0.016 }_{-0.008 }\,\mathrm{(syst.)}$ \\
  2.0 & -- & 2.5     & $0.160  \pm  0.129 \,\mathrm{(stat.)}\,^{+0.044 }_{-0.049 }\,\mathrm{(syst.)}$ \\

\hline
\end{tabular}
\end{center}
\caption{Differential cross-section
$\frac{d\sigma}{d\etamax   }$ for photons accompanied by at least one jet
in diffractive photoproduction. (Figure~\ref{fig:xsphe}(g--i))
}
\label{tab:etamax}
\end{table}

\clearpage
\newcommand{\Raplines}{ The solid histogram represents the predictions of \RAPGAP\ 
without reweighting and the dotted histogram represents
the predictions of \RAPGAP\ with reweighting, both normalised to the data. }
\newcommand{\eventcompare}{compared to
a mixture of \RAPGAP-generated direct and resolved events using the
model described in the text. 
The uncertainties shown are statistical only and no acceptance
corrections were applied at this stage.}
\newcommand{\xscomment}{The kinematic region is described in the text.
The inner error bars denote
statistical uncertainties; the outer denote  statistical with
systematic uncertainties combined in quadrature. }
\newcommand{\ncomment}{The \RAPGAP\ predictions are normalised to the data.} 

\newcommand{\ab}{\hspace*{0.05\linewidth}(a)\hspace*{0.4\linewidth}(b)}
\newcommand{\cd}{\hspace*{0.05\linewidth}(c)\hspace*{0.4\linewidth}(d)}
\newcommand{\abc}{\hspace*{0.05\linewidth}(a)\hspace*{0.3\linewidth}(b)\hspace*{0.3\linewidth}(c)}
\newcommand{\deef}{\hspace*{0.05\linewidth}(d)\hspace*{0.3\linewidth}(e)\hspace*{0.3\linewidth}(f)}
\newcommand{\ghi}{\hspace*{0.05\linewidth}(g)\hspace*{0.3\linewidth}(h)\hspace*{0.3\linewidth}(i)}
\newcommand{\jkl}{\hspace*{0.05\linewidth}(j)\hspace*{0.3\linewidth}(k)\hspace*{0.3\linewidth}(l)}

\vspace*{10mm}

%
\begin{figure}
\begin{center}
\includegraphics[width=0.45\linewidth]{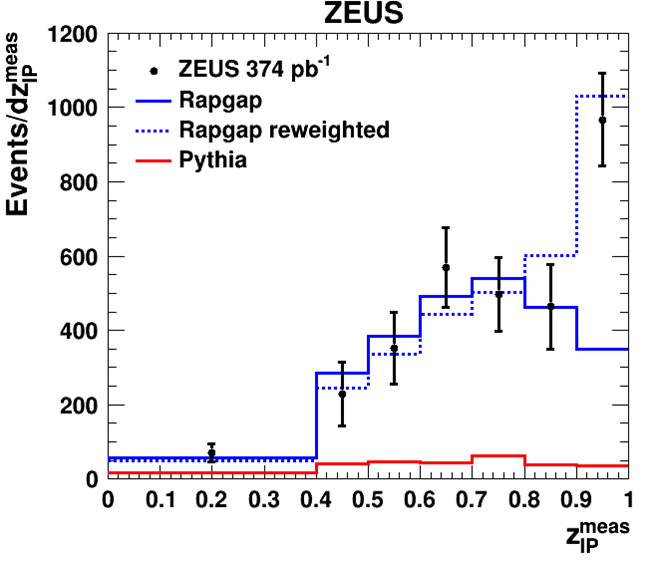}  
\includegraphics[width=0.45\linewidth]{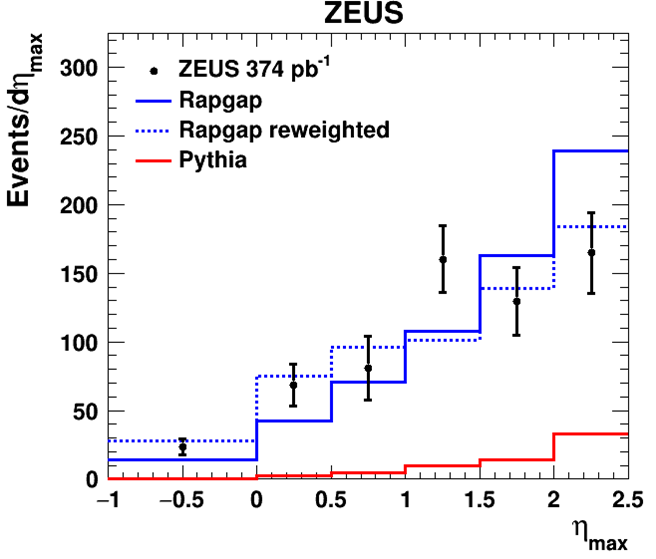}\\[-2mm] 
\ab 
\end{center}
\caption{\small  
HERA-II events with a photon and at least one jet (a) as a function of
\zpom, and (b) as a function of \etamax, per unit interval of each
variable, compared to a 70:30 \RAPGAP\ mixture of direct:resolved
photon events, with and without reweighting of the direct hadron-level
component.  The \RAPGAP histograms are normalised to the full data
sample except for the unreweighted histogram in (a), which is
normalised to the data for $\zpom < 0.9$.
 The effect of a non-diffractive contribution of 10\%, simulated with \PYTHIA,  is
indicated by the lower solid line.  }
\label{fig:rew}
\end{figure} 

\begin{figure}
\begin{center}
\includegraphics[width=0.4\linewidth]{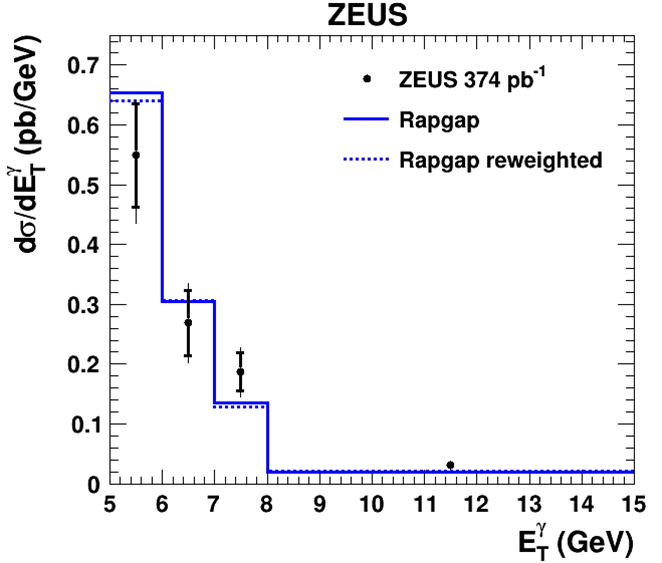}
\includegraphics[width=0.4\linewidth]{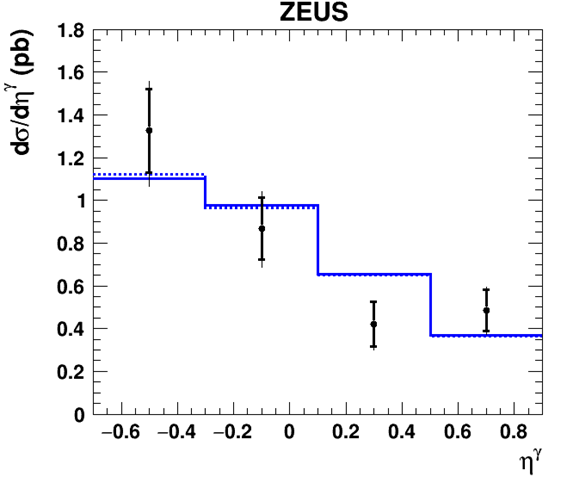}
\\[-2mm] \ab \\[4mm]
\includegraphics[width=0.4\linewidth]{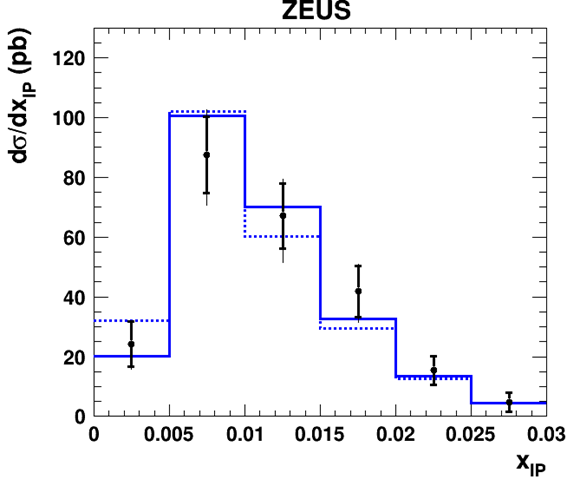}
\includegraphics[width=0.4\linewidth]{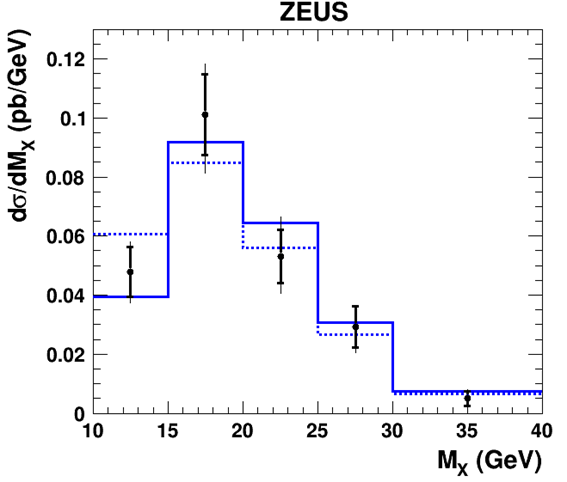}
\\[-2mm] \cd
\end{center}
\caption{\small 
Differential cross sections for inclusive isolated photon production
as functions of (a) \ETgam, (b) \etagam, (c) \xpom and (d) \Mx, measured
with HERA-II. 
\xscomment\ \ncomment\   (Tables \ref{tab:etgi}--\ref{tab:mxi})
}
\label{fig:xsincl}

\end{figure} 
%
\begin{figure}
\begin{center}
\includegraphics[width=0.7\linewidth]{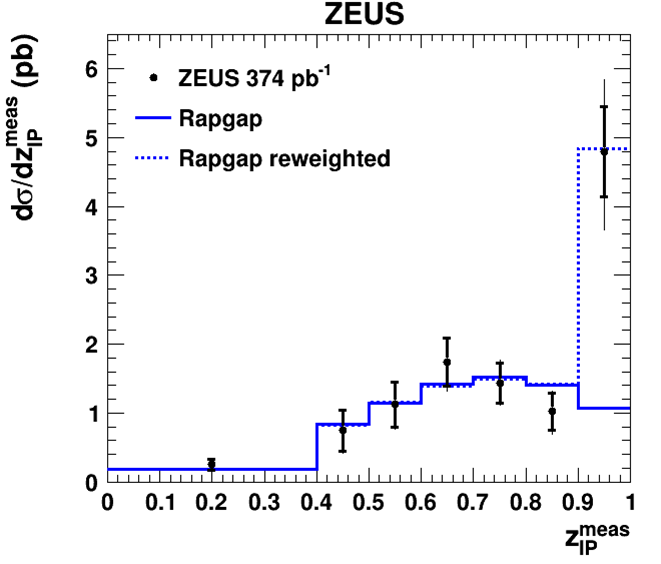}
\end{center}
\caption{\small 
Differential cross section for isolated photon production accompanied
by at least one jet, as a function of \zpom, measured with HERA-II.
The unreweighted \RAPGAP\ prediction is normalised to the data integrated over 
the region $\zpom < 0.9$; the reweighted prediction is normalised
to the full integrated data. 
\xscomment\   (Table \ref{tab:zpom})
}
\label{fig:xszpom}

\end{figure} 
%
\begin{figure}
\begin{center}
\includegraphics[width=0.3\linewidth]{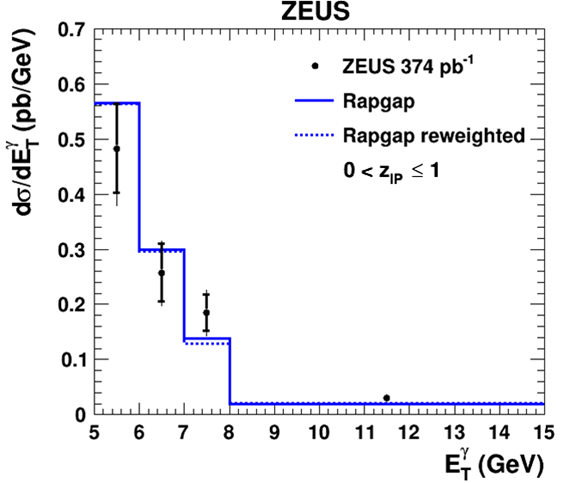}
\includegraphics[width=0.3\linewidth]{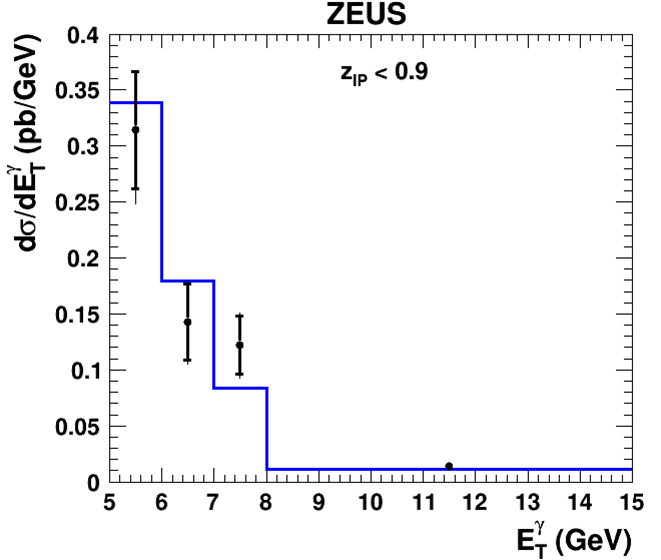}
\includegraphics[width=0.3\linewidth]{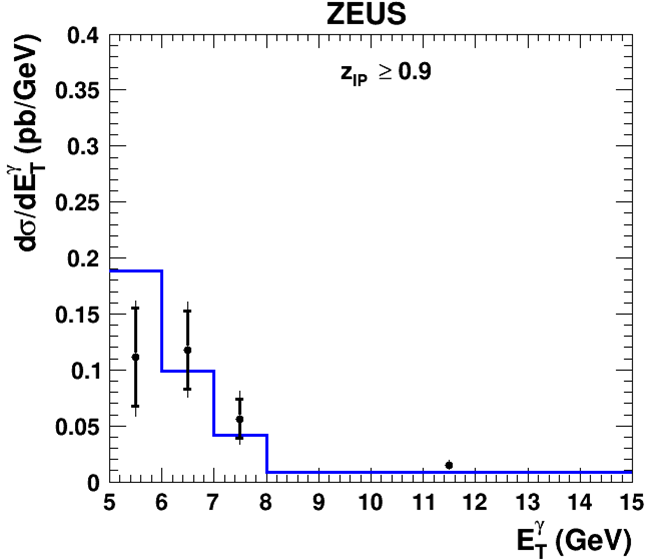}
\\[-2mm] \abc \\[4mm]
\includegraphics[width=0.3\linewidth]{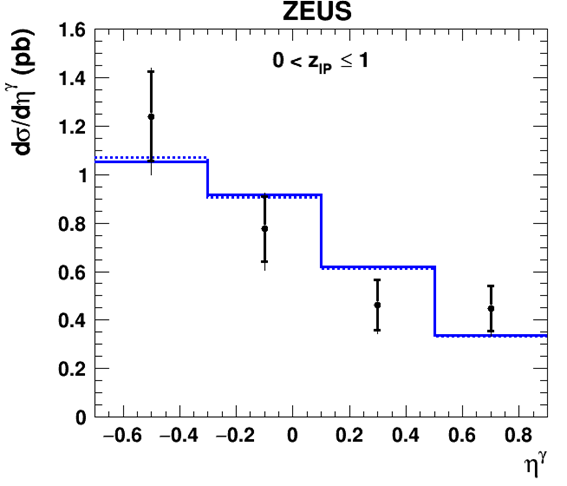}
\includegraphics[width=0.3\linewidth]{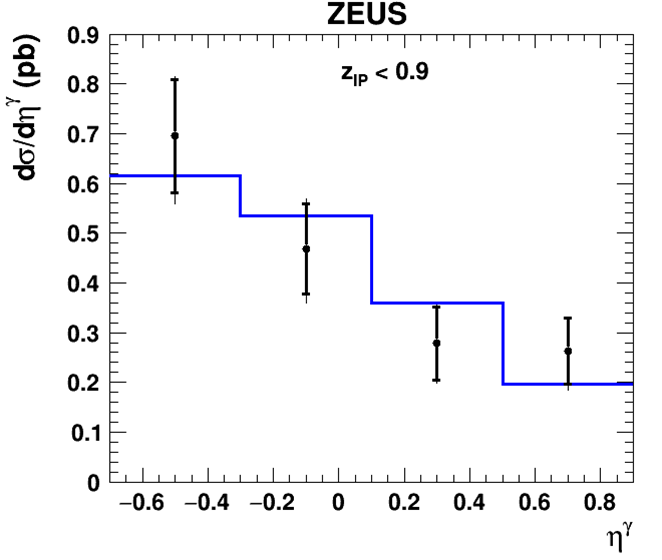}
\includegraphics[width=0.3\linewidth]{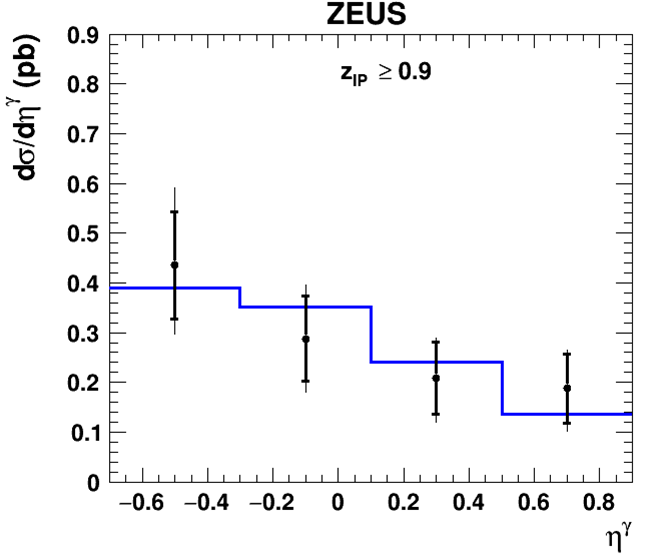}
\\[-2mm] \deef \\[4mm]
\includegraphics[width=0.3\linewidth]{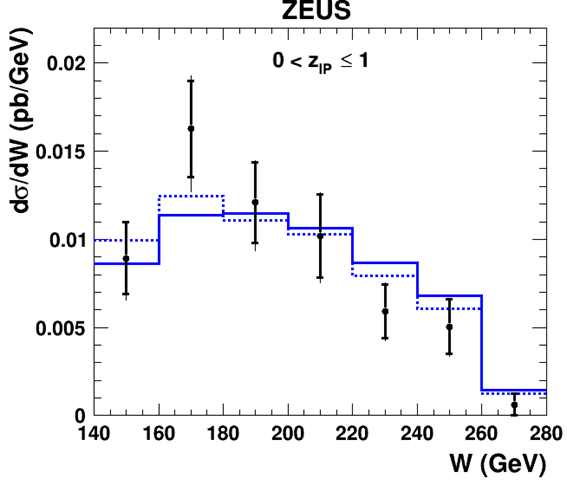}
\includegraphics[width=0.3\linewidth]{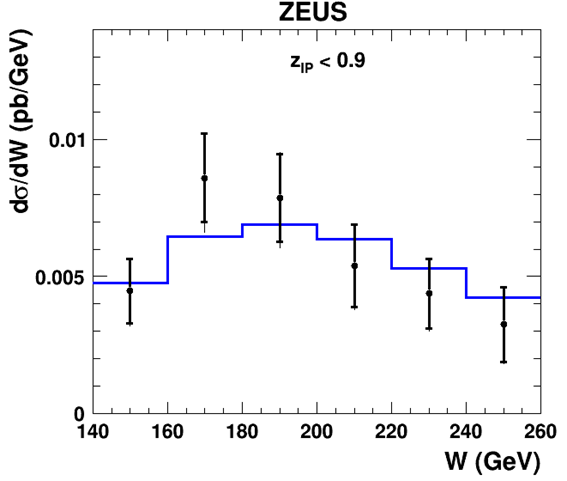}
\includegraphics[width=0.3\linewidth]{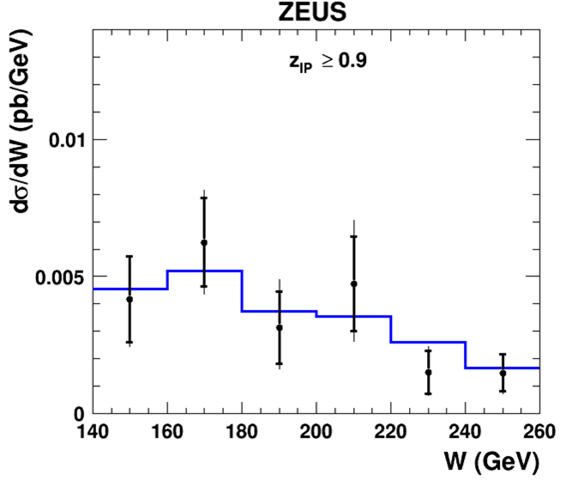}
\\[-2mm] \ghi \\[4mm]
\end{center}
\caption{\small  
Differential cross sections for isolated photon production accompanied
by at least one jet, as functions of (a--c) \ETgam, (d--f) \etagam, and (g--i)
$W$, measured with HERA-II.  Results are presented for (a, d, g) using
the full \zpom\ range, (b, e, h) $\zpom\ < 0.9$, and (c, f, i) $\zpom
\geq 0.9$. The \RAPGAP\ predictions are normalised to the data in the
selected range; the reweighted prediction is shown in (a, d, g) only
since in the other plots the normalisation makes the two predictions
identical.
\xscomment\  (Tables \ref{tab:etgj}--\ref{tab:Wj}) 
 }
\label{fig:xsgam}
\end{figure} 

\begin{figure}
\begin{center}
\includegraphics[width=0.3\linewidth]{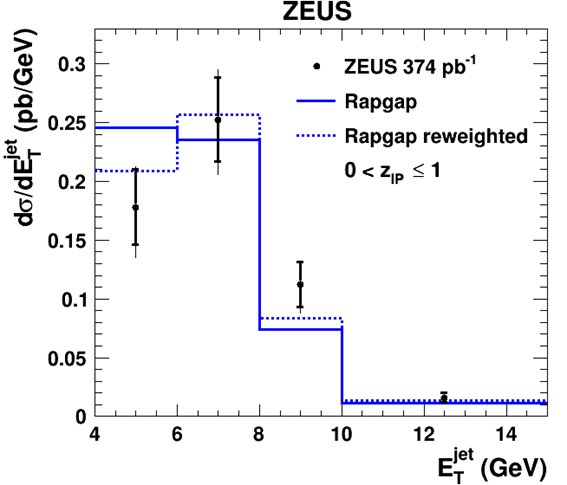}
\includegraphics[width=0.3\linewidth]{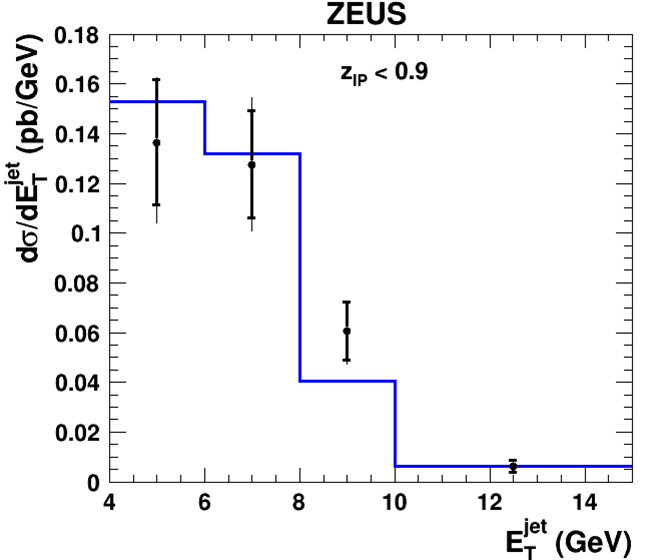}
\includegraphics[width=0.3\linewidth]{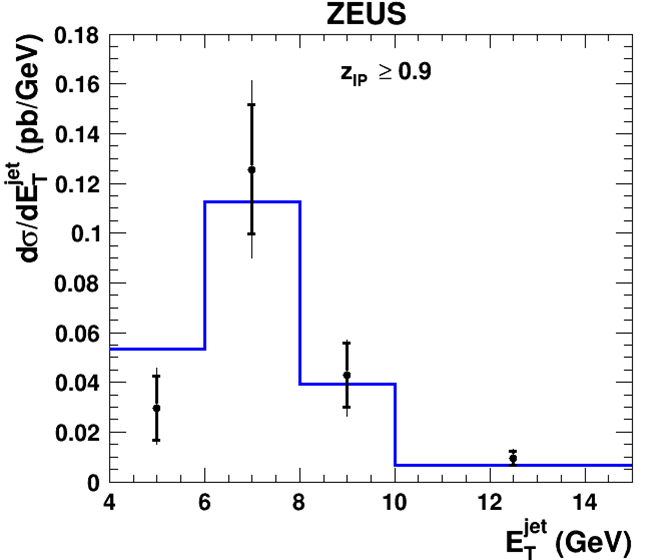}
\\[-2mm] \abc \\[4mm]
\includegraphics[width=0.3\linewidth]{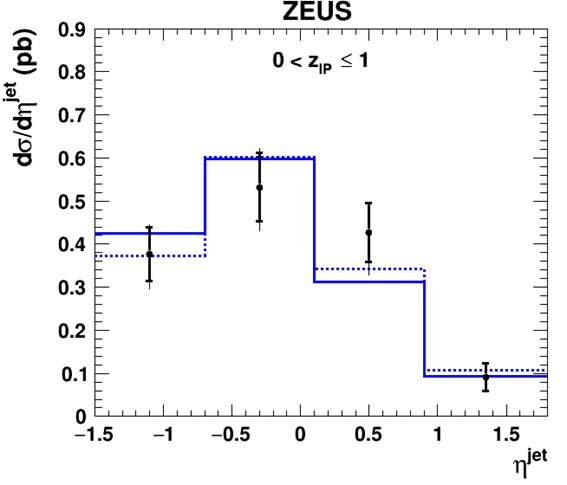}
\includegraphics[width=0.3\linewidth]{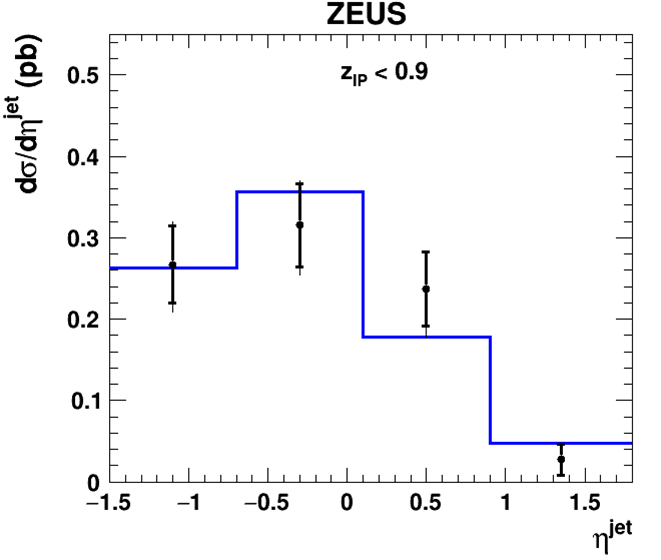}
\includegraphics[width=0.3\linewidth]{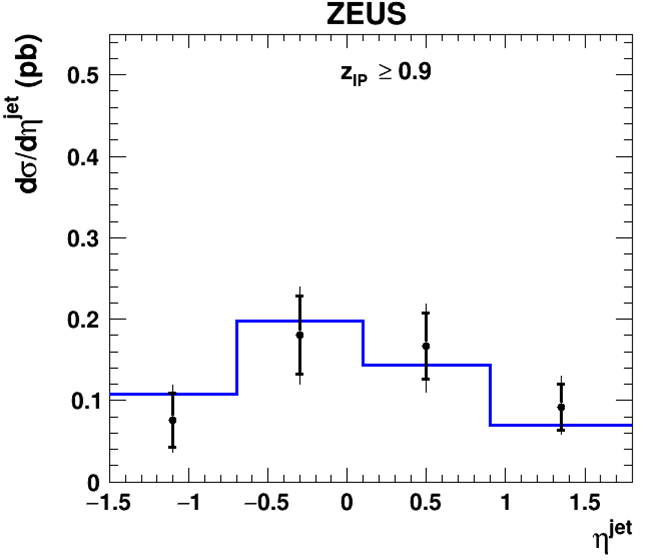}
\\[-2mm] \deef \\[4mm]
\includegraphics[width=0.3\linewidth]{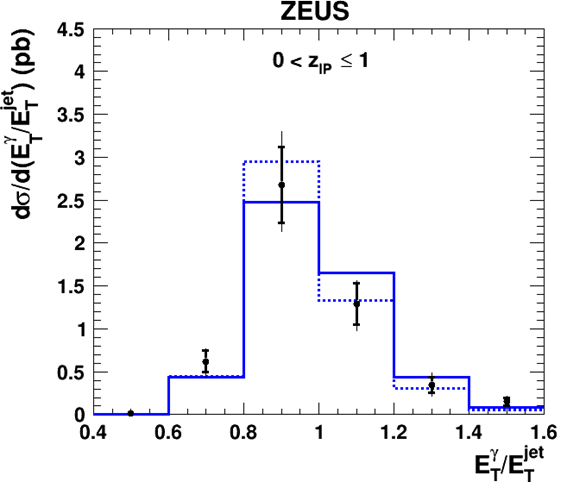}
\includegraphics[width=0.3\linewidth]{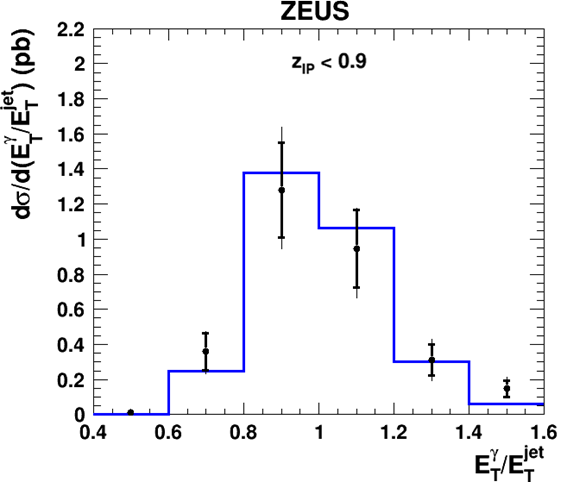}
\includegraphics[width=0.3\linewidth]{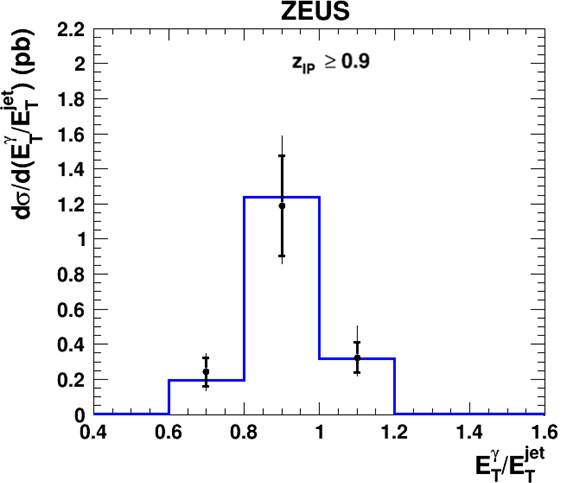}
\\[-2mm] \ghi \\[4mm]
\end{center}
\caption{\small  
Differential cross sections for isolated photon production accompanied
by at least one jet, as functions of (a--c) \ETjet, (d--f)
\etajet, and (g--i) the transverse energy ratio \ETgam/\ETjet measured with HERA-II. 
Results are presented for (a, d, g) the full
\zpom\ range, (b, e, h) $\zpom\ < 0.9$, and (c, f, i) $\zpom \geq 0.9$.
 Other details as in Fig.~\ref{fig:xsgam}. 
 (Tables \ref{tab:etjj}--\ref{tab:etrat})
 }
\label{fig:xsjet}

\end{figure} 

\begin{figure}
\begin{center}
\includegraphics[width=0.3\linewidth]{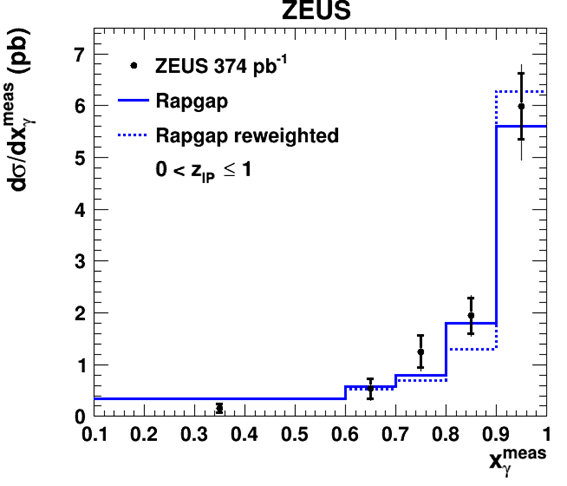}
\includegraphics[width=0.3\linewidth]{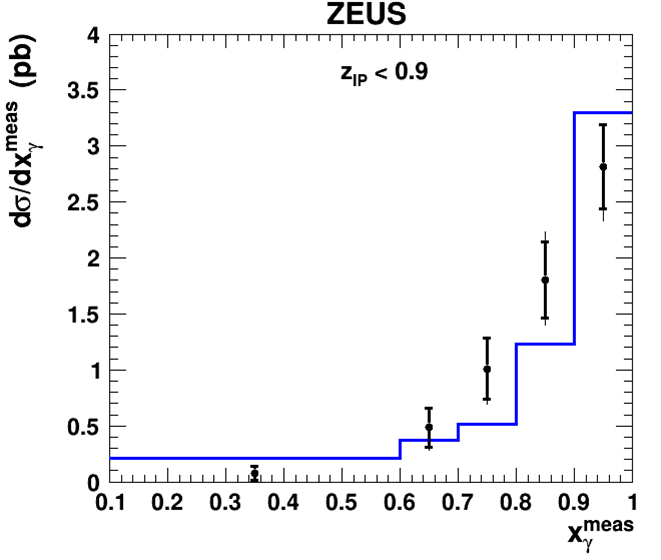}
\includegraphics[width=0.3\linewidth]{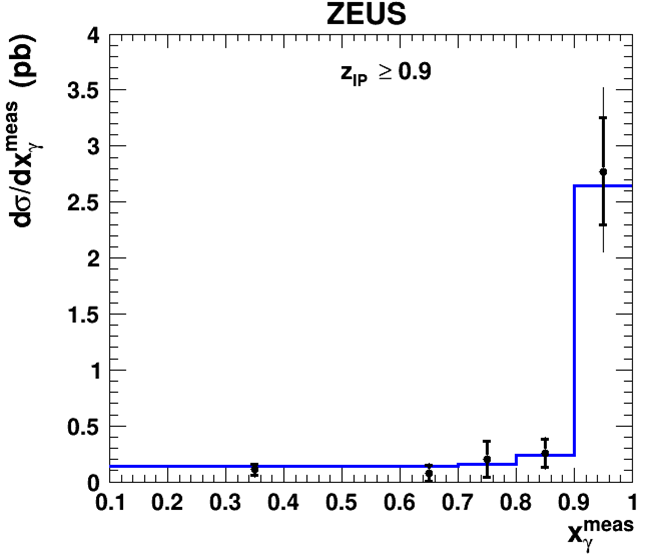}
\\[-2mm] \abc \\[4mm]
\includegraphics[width=0.3\linewidth]{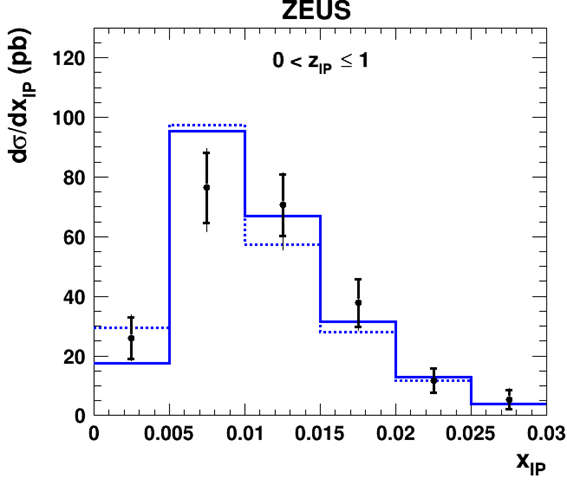}
\includegraphics[width=0.3\linewidth]{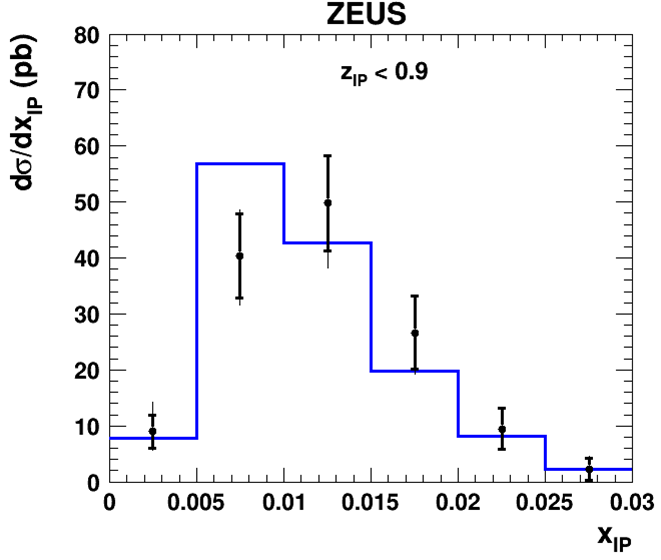}
\includegraphics[width=0.3\linewidth]{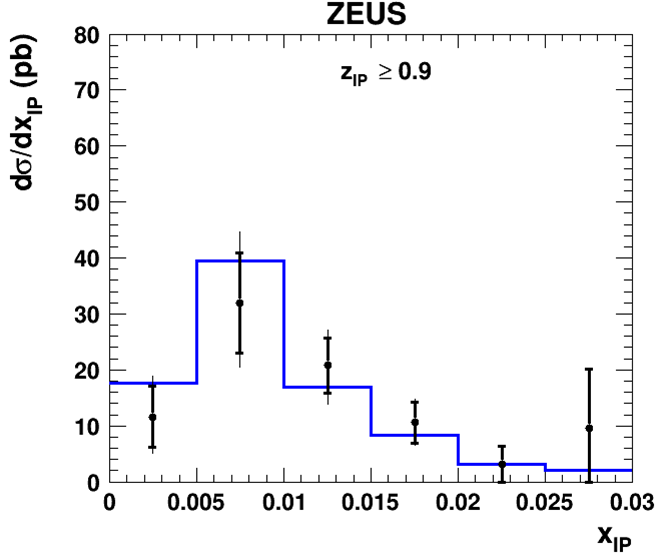}
\\[-2mm] \deef \\[4mm]
\includegraphics[width=0.3\linewidth]{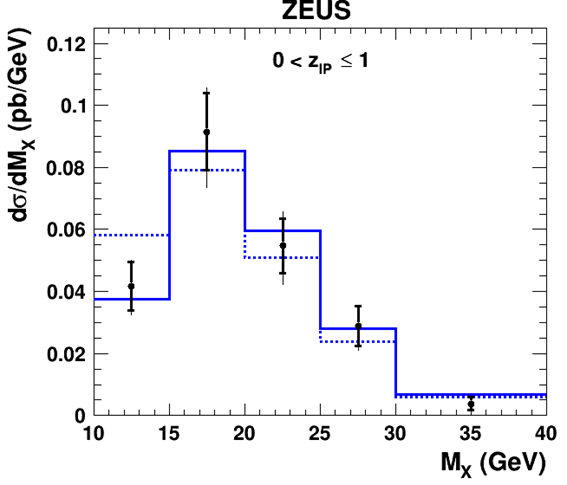}
\includegraphics[width=0.3\linewidth]{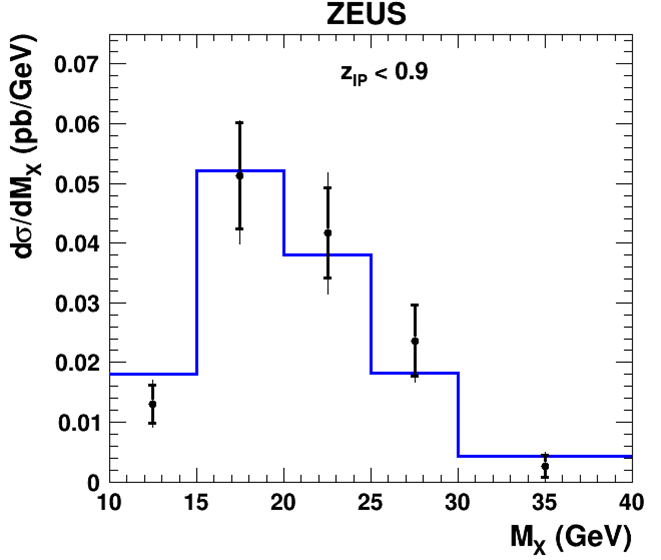}
\includegraphics[width=0.3\linewidth]{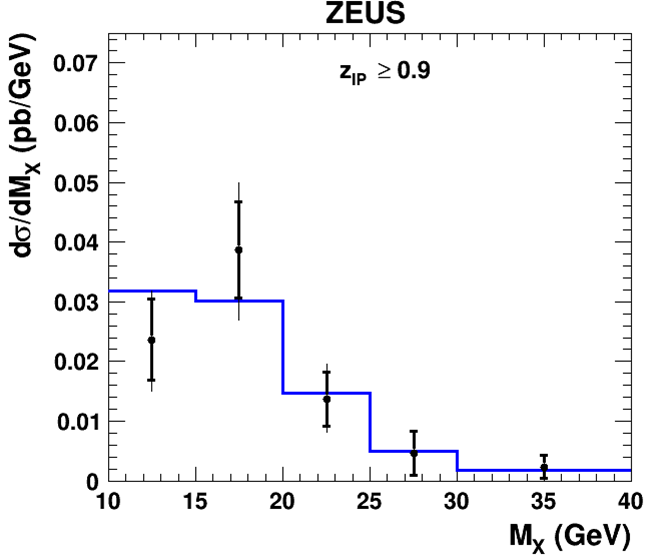}
\\[-2mm] \ghi \\[4mm]
\end{center}
\caption{\small  
Differential cross sections for isolated photon production accompanied
by at least one jet, as functions of (a--c) \xgamm\, (d--f) \xpom, and (g--i) \Mx,
measured with HERA-II.  Results are presented for (a, d, g) the full
\zpom\ range, (b, e, h) $\zpom\ < 0.9$, and (c, f, i) $\zpom \geq 0.9$.
Other details as in Fig.~\ref{fig:xsgam}.
 (Tables \ref{tab:xg}--\ref{tab:mxj})
 }

\label{fig:xsxm}

\end{figure}

\newpage

\begin{figure}
\begin{center}
~\\[-10mm]
\includegraphics[width=0.3\linewidth]{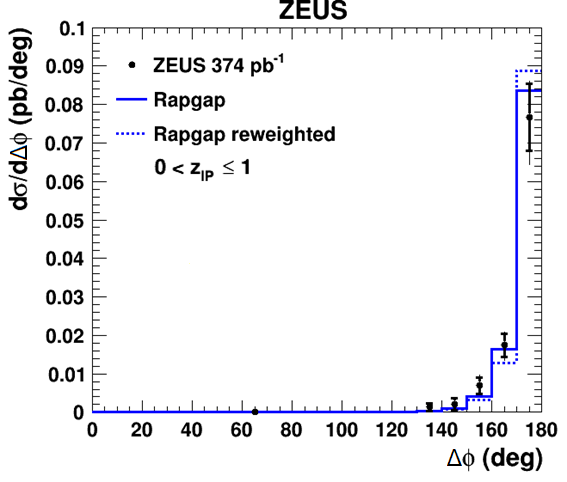}
\includegraphics[width=0.3\linewidth]{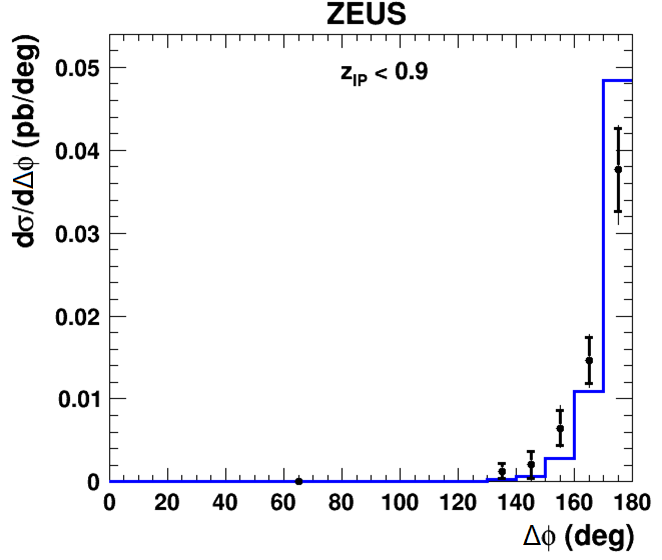}
\includegraphics[width=0.3\linewidth]{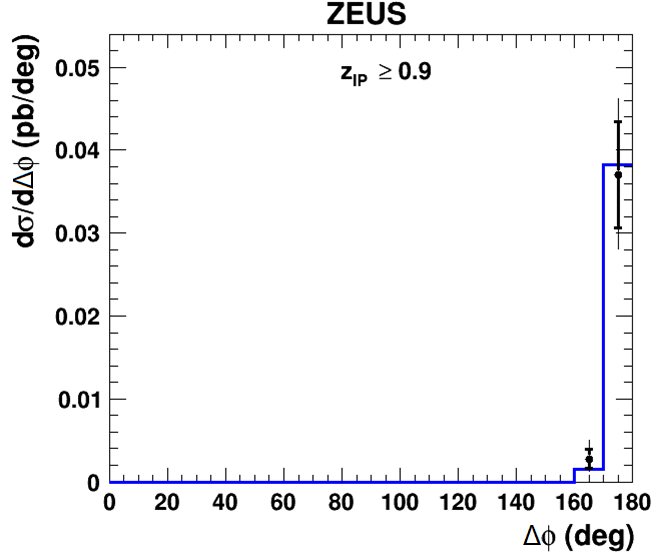}
\\[-2mm] \abc \\[4mm]
\includegraphics[width=0.3\linewidth]{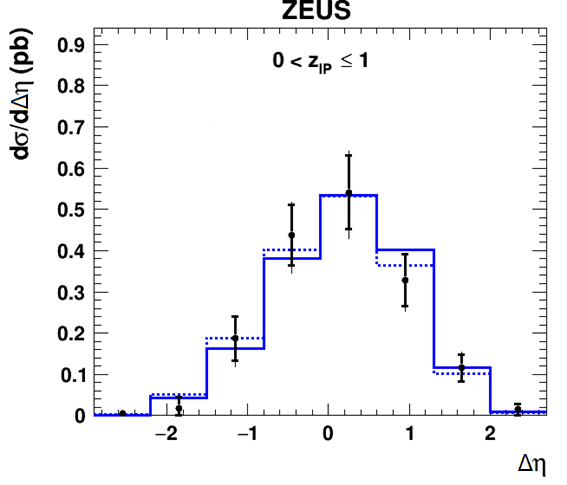}
\includegraphics[width=0.3\linewidth]{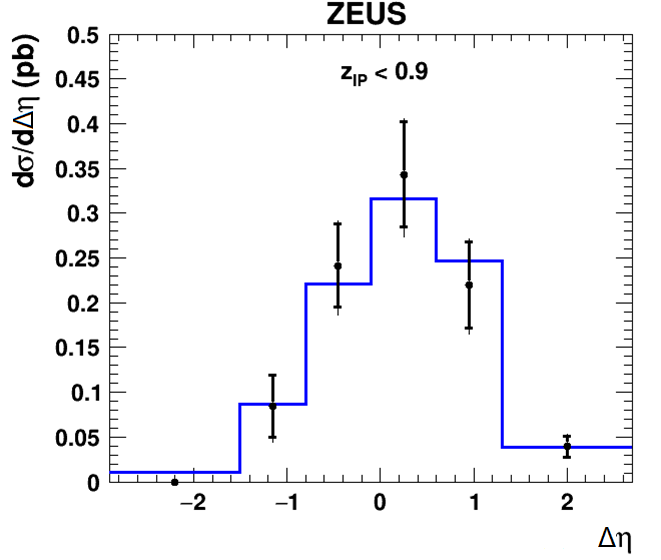}
\includegraphics[width=0.3\linewidth]{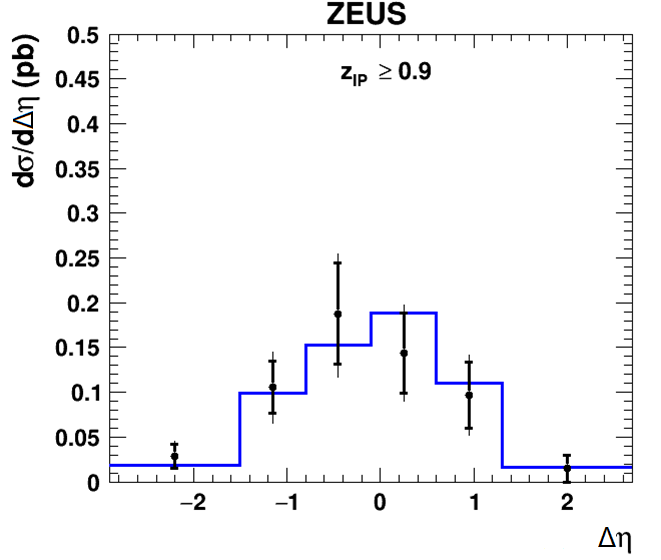}
\\[-2mm] \deef \\[4mm]
\includegraphics[width=0.3\linewidth]{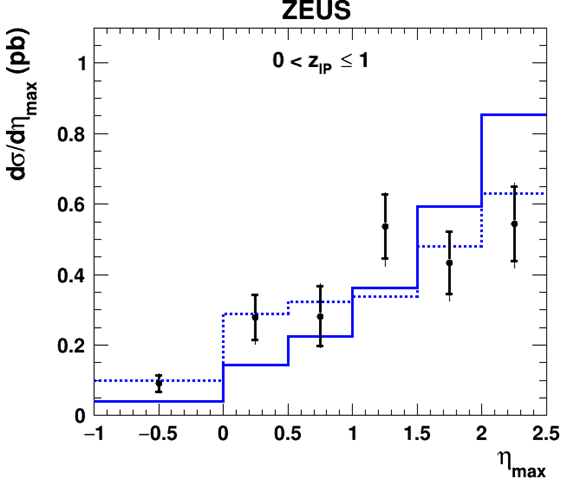}
\includegraphics[width=0.3\linewidth]{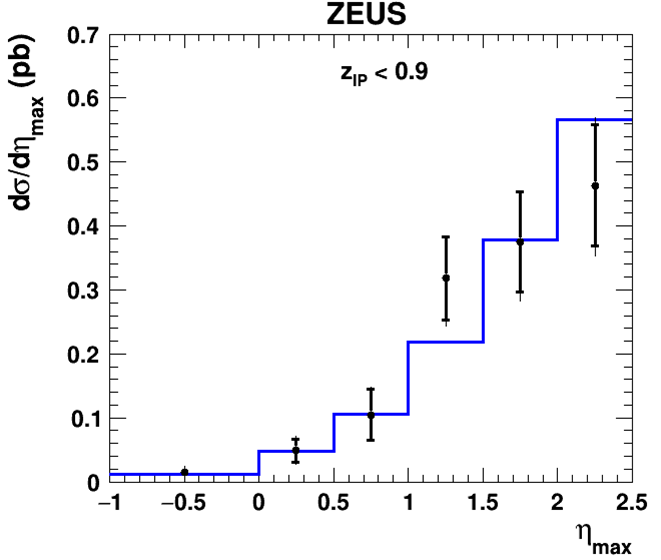}
\includegraphics[width=0.3\linewidth]{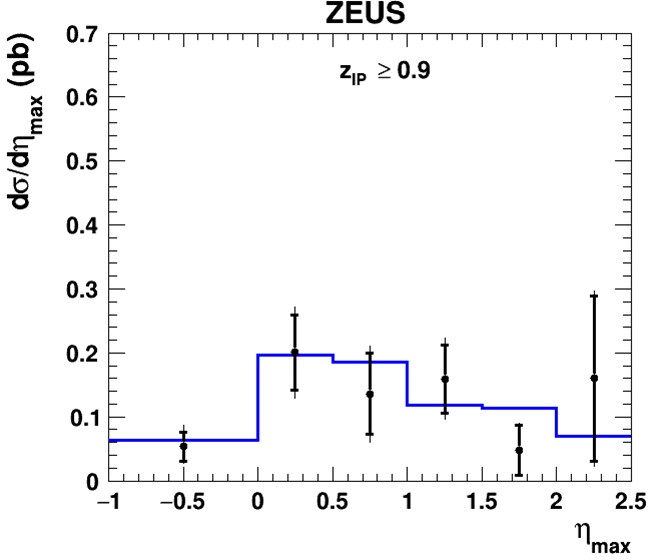}
\\[-2mm] \ghi \\[4mm]
\end{center}
\caption{\small  
Differential cross sections for isolated photon production accompanied
by at least one jet, as functions of (a--c) $\delphi = |\phi^\gamma -
\phi^{\mathrm jet}|$, (d--f) $\Deleta = \eta^\gamma - \etajet$, 
and (g--i) \etamax, measured with HERA-II. Results are presented for (a, d, g) the full
\zpom\ range, (b, e, h) $\zpom\ < 0.9$, and (c, f, i) $\zpom \geq 0.9$.
Other details as in Fig.~\ref{fig:xsgam}. 
(Tables \ref{tab:phi}--\ref{tab:etamax})
 }
\label{fig:xsphe}
\end{figure}

%

%

\begin{mcbibliography}{10}
\bibitem{mrw}
G. Watt, A.D. Martin and M.G. Ryskin,
\mbox{\tt arXiv:0708.4126};\\
A. Martin, M. Ryskin and G. Watt,
\newblock Phys.\ Lett.\ B 644 (2007) 131\relax
\relax
\bibitem{twogluons}
M. W\"usthoff, \newblock  Ph.D Thesis, DESY-95-166 (1995);\\
M. Diehl, \newblock Z. Phys.\ C76 (1997) 499;\\
J. Bartels \etal, in Proc.\ {\it Workshop on Future Physics at HERA,}
eds. A. De Roeck, G. Ingelman and R. Klanner,  DESY (1995/966) 668,  
\mbox{\tt hep-ph/9609239}\relax
\relax
\bibitem{rapgap}
H.~Jung, \mbox{\tt https://rapgap.hepforge.org/rapgap.pdf}\relax
\relax
\bibitem{ingelman}
G. Ingelman and P. Schlein, 
\newblock Phys.\ Lett.{} B~152~(1985) 256\relax
\relax
\bibitem{epjc:48:715}
H1~\coll, A. Aktas \etal, 
\newblock Eur.\ Phys.\ J. C~48~(2006) 715\relax
\relax
\bibitem{zp:c65:657}
B. Kniehl, H.-G.\ Kohrs and G. Kramer, 
\newblock Z.\ Phys.\ C 65 (1995) 657;\\
H.-G.\ Kohrs, \mbox{\tt hep-ph/9507208};\\
A. Donnachie and P.V. Landshoff,
\newblock Phys.\ Lett.\ B 285 (1992) 172\relax
\relax
\bibitem{Collins}
J. Collins, \newblock Phys.\ Rev.\ D 57 (1998) 3051,
erratum Phys.\ Rev.\ D 61 (2000) 019902\relax
\relax
\bibitem{epj:c21:521}
A. Kaidalov \etal, 
\newblock Eur.\ Phys.\ J. C 21 (2001) 521\relax
\relax
\bibitem{kandk}
M. Klasen and G. Kramer, 
in Proc.\ {\it 12 International Workshop on Deep Inelastic Scattering,} 
eds.\ D. Bruncko, J. Ferencei and P. Stri\v{z}ene\'{c} (2004) 492, 
 \mbox{\tt hep-ph/0401202}\relax
\relax
\bibitem{epj:c70:15}
H1 Collaboration, F.D. Aaron \etal,
\newblock Eur.\ Phys.\ J. C 70 (2010) 15\relax
\relax
\bibitem{epj:c6:421}
H1 Collaboration, C. Adloff \etal,
\newblock Eur.\ Phys.\ J. C 6 (1999) 421\relax
\relax
\bibitem{epj:c55:177}
ZEUS Collaboration, S. Chekanov \etal,
\newblock Eur.\ Phys.\ J. C 55 (2008) 177\relax
\relax
\bibitem{jhep:05:056}
H1 Collaboration, V. Andreev \etal, 
\newblock JHEP 1505 (2015) 056\relax
\relax
\bibitem{epj:c5:41}
ZEUS \coll, J. Breitweg \etal, 
\newblock Eur.\ Phys.\ J. C 5 (1998) 41\relax
\relax 
\bibitem{epj:c51:549}
H1 \coll, A. Aktas \etal, 
\newblock Eur.\ Phys.\ J. C 51 (2007) 549\relax 
\relax
\bibitem{plb:672:219}
H1 Collaboration, V. Andreev \etal,
\newblock Phys.\ Lett.\ B 672 (2009) 219\relax
\relax
\bibitem{pl:b730:293}
ZEUS \coll, H. Abramowicz \etal,
 Phys.\ Lett.\ B~730~(2014)~293\relax 
\relax
\bibitem{pl:b413:201}
ZEUS \coll, J.~Breitweg \etal,
\newblock Phys.\ Lett.{} B~413~(1997)~201\relax
\relax
\bibitem{pl:b472:175}
ZEUS \coll, J.~Breitweg \etal,
\newblock Phys.\ Lett.{} B~472~(2000)~175\relax
\relax
\bibitem{pl:b511:19}
ZEUS \coll, S.~Chekanov \etal,
\newblock Phys.\ Lett.{} B~511~(2001)~19\relax
\relax
\bibitem{epj:c49:511}
ZEUS \coll, S. Chekanov \etal,
\newblock Eur.\ Phys.\ J.{} C~49~(2007)~511\relax
\relax
\bibitem{epj:c38:437}
H1 \coll, A.~Aktas \etal,
\newblock Eur.\ Phys.\ J.{} C~38~(2004)~437\relax
\relax
\bibitem{epj:c66:17}
H1 \coll, F.D.~Aaron \etal, Eur.\ Phys.\ J.{} C~66~(2010)~17\relax
\bibitem{epj:c54:371}
H1 \coll, F.D.~Aaron \etal,
\newblock Eur.\ Phys.\ J.{} C~54~(2008)~371\relax
\relax
\bibitem{pl:b595:86}
ZEUS \coll, S.~Chekanov \etal,
\newblock Phys.\ Lett.{} B~595~(2004)~86\relax
\relax
\bibitem{pl:b687:16}
ZEUS \coll, S.~Chekanov \etal,
\newblock Phys.\ Lett.{} B~687~(2010)~16\relax
\relax
\bibitem{pl:b715:88}
ZEUS \coll, H. Abramowicz \etal,
\newblock Phys.\ Lett.{} B~715~(2012)~88\relax
\relax
\bibitem{zeus:1993:bluebook}
ZEUS \coll, U.~Holm~(ed.),
\newblock {\em The {ZEUS} Detector}.
\newblock Status Report (unpublished), DESY (1993),
\newblock available on
  \texttt{http://www-zeus.desy.de/bluebook/bluebook.html}\relax
\relax
\bibitem{nim:a279:290}
N.~Harnew \etal,
\newblock Nucl.\ Inst.\ Meth.{} A~279~(1989)~290\relax
\relax
\bibitem{npps:b32:181}
B.~Foster \etal,
\newblock Nucl.\ Phys.\ Proc.\ Suppl.{} B~32~(1993)~181\relax
\relax
\bibitem{nim:a338:254}
B.~Foster \etal,
\newblock Nucl.\ Inst.\ Meth.{} A~338~(1994)~254\relax
\relax
\bibitem{nim:a581:656}
A.~Polini \etal,
\newblock Nucl.\ Inst.\ Meth.{} A~581~(2007)~656\relax
\relax
\bibitem{nim:a309:77}
M.~Derrick \etal,
\newblock Nucl.\ Inst.\ Meth.{} A~309~(1991)~77\relax
\relax
\bibitem{nim:a309:101}
A.~Andresen \etal,
\newblock Nucl.\ Inst.\ Meth.{} A~309~(1991)~101\relax
\relax
\bibitem{nim:a321:356}
A.~Caldwell \etal,
\newblock Nucl.\ Inst.\ Meth.{} A~321~(1992)~356\relax
\relax
\bibitem{nim:a336:23}
A.~Bernstein \etal,
\newblock Nucl.\ Inst.\ Meth.{} A~336~(1993)~23\relax
\relax
\bibitem{fpc}
A. Bamburger \etal,
\newblock Nucl.\ Inst.\ Meth.{} A~450~(2000)~235\relax
\relax
\bibitem{lumi2}
L.~Adamczyk \etal, 
\newblock Nucl.\ Inst.\ Meth.{} A~744~(2014)~80\relax
\relax
\bibitem{desy-92-066}
J.~Andruszk\'ow \etal,
\newblock Preprint \mbox{DESY-92-066}, DESY, 1992\relax
\relax
\bibitem{zfp:c63:391}
ZEUS \coll, M.~Derrick \etal,
\newblock Z.\ Phys.{} C~63~(1994)~391\relax
\relax
\bibitem{acpp:b32:2025}
J.~Andruszk\'ow \etal,
\newblock Acta Phys.\ Pol.{} B~32~(2001)~2025\relax
\relax
\bibitem{nim:a565:572}
M.~Helbich \etal,
\newblock Nucl.\ Inst.\ Meth.{} A~565~(2006)~572\relax
\relax
\bibitem{epj:c76:1}
ZEUS \coll, H. Abramowicz \etal, 
\newblock Eur.\ Phys.\ J. C~76~(2016)~1\relax
\relax 
\bibitem{yjb}
F. Jacquet and A. Blondel, 
\newblock Proc. {\it Study for an ep Facility for Europe,} 
DESY 79/48 (1979) 391\relax
\relax
\bibitem{pl:b303:183}
ZEUS \coll, M.~Derrick \etal,
\newblock Phys.\ Lett.{} B~303~(1993)~183\relax
\relax
\bibitem{tech:cern-dd-ee-84-1}
R.~Brun \etal, {\em {\sc Geant3}},
\newblock Technical Report CERN-DD/EE/84-1, CERN (1987)\relax
\relax
\bibitem{jung}
H.~Jung, private communication\relax
\relax
\bibitem{jhep:0605:026}
T.~Sj\"ostrand \etal,
\newblock JHEP{} 05~(2006)~26\relax
\relax
\bibitem{plb:376:193}
G. Schuler and T. Sj\"ostrand, 
\newblock Phys.\ Lett.\ B~376~(1996)~193\relax
\relax
\bibitem{pr:d55:1280}
H.L. Lai \etal, \relax
\newblock Phys.\ Rev. D 55 (1997) 1280\relax
\relax
\bibitem{grv}
M. Gl\"uck, G. Reya and A. Vogt,
\newblock Phys.\ Rev.\ D~45~(1992)~3986;
\\
M. Gl\"uck, G. Reya and A. Vogt,
\newblock Phys.\ Rev.\ D~46~(1992) 1973\relax
\relax
\bibitem{grape}
T. Abe, \relax
\newblock Comp.\ Phys.\ Comm.\ 136 (2001) 126\relax
\relax
\bibitem{djangoh}
H. Spiesberger, {\it HERACLES and DJANGOH Event Generators for ep Interactions at HERA 
Including Radiative Processes,} 1998,  
\mbox{\tt http://wwwthep.physik.uni-mainz.de/\raisebox{-0.6ex}{\~{}}hspiesb/djangoh/djangoh.html}\relax
\relax
\bibitem{ariadne}
L. L\"onnblad,
\newblock Comp.\ Phys.\ Comm.\ 71 (1992) 15;\\
L. L\"onnblad,   
\newblock  Z. Phys.\ C 65 (1995) 285\relax
\relax
\bibitem{uproc:chep:1992:222}
W.H.~Smith, K.~Tokushuku and L.W.~Wiggers,
\newblock {\em Proc.\ Computing in High-Energy Physics (CHEP), Annecy, France,
  Sep.\ 1992}, eds. C.~Verkerk and W.~Wojcik,
\newblock CERN, Geneva, Switzerland (1992) 222.
\newblock Also in preprint \mbox{DESY 92-150B}\relax
\relax
\bibitem{nim:a580:1257}
P.~Allfrey \etal,
\newblock Nucl.\ Inst.\ Meth.{} A~580~(2007)~1257\relax
\relax
\bibitem{epj:c1:81}
ZEUS \coll, J.~Breitweg \etal,
\newblock Eur.\ Phys.\ J.{} C~1~(1998)~81\relax
\relax
\bibitem{epj:c6:43}
ZEUS \coll, J.~Breitweg \etal,
\newblock Eur.\ Phys.\ J.{} C~6~(1999)~43\relax
\relax
\bibitem{andrii}
A. Iudin, Ph.D. Thesis, 
\newblock Kyiv National University ``Kyiv Polytechnic Institute'' (2014, unpublished)\relax
\relax
\bibitem{np:b406:187}
S.~Catani \etal,
\newblock Nucl.\ Phys.{} B~406~(1993)~187\relax
\relax
\bibitem{pr:d48:3160}
S.D.~Ellis and D.E.~Soper,
\newblock Phys.\ Rev.{} D~48~(1993)~3160\relax
\relax
\bibitem{kagawa}
S. Kagawa, Ph.D.\ thesis, KEK Report 2005-12 (2006)\relax
\relax
\bibitem{unfold}
G. D'Agostini, 
\newblock Nucl.\ Instr.\ Meth.{} A~362~(1995)~487;\\
G. Kondor, 
\newblock Nucl.\ Instr.\ Meth.{} A~216~(1983)~177;\\
H. M\"ulthei and B. Schorr,
\newblock Nucl.\ Instr.\ Meth.{} A~257~(1987)~371\relax
\relax
\end{mcbibliography}
\end{document}